\documentclass[useAMS,usenatbib]{mn2e}
\usepackage{graphicx}
\usepackage{times}

\title[Momentum transfer in SPH simulations]{Momentum transfer across shear flows in Smoothed Particle Hydrodynamic simulations of galaxy formation}
\author[T. Okamoto et al.]{T. Okamoto$^{1}$\thanks{E-mail:
Takashi.Okamoto@durham.ac.uk (TO); A.R.Jenkins@durham.ac.uk (AJ);
V.R.Eke@durham.ac.uk (VRE); Vicent.Quilis@uv.es (VQ); C.S.Frenk@durham.ac.uk (CSF)},
A. Jenkins$^{1}$, V. R. Eke$^{1}$, V. Quilis$^{2}$, and C. S. Frenk$^{1}$ \\
$^{1}$Institute for Computational Cosmology, Physics Department, Durham University, South Road, Durham DH1 3LE,
UK\\
$^{2}$Departament d'Astronomia i Astrof\'{\i}sica, Universitat de Val`encia,
       46100-Burjassot(Valencia), Spain
}
\begin{document}

\date{}

\pagerange{\pageref{firstpage}--\pageref{lastpage}} \pubyear{2003}

\maketitle

\label{firstpage}

\begin{abstract}
We investigate the evolution of angular momentum in Smoothed Particle
Hydrodynamic (SPH) simulations of galaxy formation, paying particular
attention to artificial numerical effects.  We find that a cold gas
disc forming in an ambient hot gas halo receives a strong hydrodynamic
torque from the hot gas.  By splitting the hydrodynamic force into
artificial viscosity and pressure gradients, we find that the angular
momentum transport is caused not by the artificial viscosity but by
the pressure gradients.  Using simple test simulations of shear flows,
we conclude that the pressure gradient-based viscosity can be divided
into two components: one due to the noisiness of SPH and the other to
ram pressure.  The former is problematic even with very high
resolution because increasing resolution does not reduce the
noisiness. On the other hand, the ram pressure effect appears only
when a cold gas disc or sheet does not contain enough particles. In
such a case, holes form in the disc or sheet, and then ram pressure
from intra-hole hot gas, causes significant deceleration.  In
simulations of galactic disc formation, star formation usually
decreases the number of cold gas particles, and hole formation leads
to the fragmentation of the disc. This fragmentation not only induces
further angular momentum transport, but also affects star formation in
the disc.  To circumvent these problem, we modify the SPH algorithm,
decoupling the cold from the hot gas phases, i.e. inhibiting the
hydrodynamic interaction between cold and hot particles.  This, a
crude modelling of a multi-phase fluid in SPH cosmological
simulations, leads to the formation of smooth extended cold gas discs
and to better numerical convergence.  The decoupling is applicable in
so far as the self-gravitating gas disc with negligible external
pressure is a good approximation for a cold gas disc.

\end{abstract}


\section{INTRODUCTION}

Understanding the formation of galactic discs is one of the most
important, unsolved problems in astrophysics.  In the currently
favoured cold dark matter (CDM) cosmological framework, in which
structure builds up hierarchically \citep{defw85}, discs are assumed
to form in the potential wells of virialised dark matter halos
through radiative cooling \citep{wr78, fe80, dss97, mmw98, vdb01}.  In
this scenario, baryons are required to retain most of the angular
momentum imparted to them by tidal torques in order for the resulting
centrifugally supported discs to have realistic sizes.  The
conservation of gas angular momentum is an important assumption made
in semi-analytic modelling of disc formation
\citep{kau93, sp99, col00, nag01, on03}.

However, to date, numerical simulations of galaxy formation starting
from appropriate CDM initial conditions, and allowing just radiative
cooling of the gas and no star formation or feedback, find that the
infalling gas loses too much angular momentum.  This results in discs
forming which are much too small
\citep[e.g.][]{nw94, nfw95}.  This problem is commonly called ``the
angular momentum problem''.  The angular momentum losses arise during
the hierarchical clustering process. At early times in a CDM dominated
universe, small dense dark matter halos form. Radiative cooling is
very efficient in these objects and a large fraction of the gas cools into
them. As these gas rich halos merge to form larger halos their
incoming orbital angular momentum is drained by dynamical friction and
exported to the dark matter at the outskirts of the new halos.  Much
of the original angular momentum of the gas is lost through these
processes by the time it reaches the middle and forms a disc.

\citet{wee98} and \citet{eew00} have shown that if cooling is
suppressed until the host halos are well established, then the
numerical simulations yield much larger discs. Two ideas have been
suggested which might prevent the early collapse of small
proto-galactic gas clouds.  One is that cooling may be suppressed by
feedback due to energy injected by stellar winds and supernovae.
Simulations invoking very energetic feedback have illustrated the
possibility of resolving the angular momentum problem in this way.
\citep{tc01,sgp02}. The second idea that has been suggested to prevent
the formation of small proto-galactic clouds is to invoke an
alternative form of dark matter, ``warm dark matter,'' in which
case the initial density field does not have small scale fluctuations
\citep{pp82}. \citet{sd01} and \citet{gov02} have shown that galaxies
formed in this model have larger discs and smaller bulges than in
simulations with CDM.  The angular momentum problem is potentially a
strong clue which can help unravel the processes of galaxy formation
and the complicated star formation and feedback processes
involved. For this to be possible, however, we must be careful to
understand the role of any numerical effects which may be important in
determining the outcome of galaxy formation simulations.

Smoothed particle hydrodynamics (SPH) has been widely used to study
galaxy formation \citep[e.g. ][]{khw92, esd94, nfw95, sn99, tc00,
sn02} both because its fully Lagrangian nature is suited to problems
that need a wide dynamic range like galaxy formation, and because of
its simplicity and robustness which make it easy to incorporate into
$N$-body codes.  Despite these attractive features, there are
problems.  First of all, most of the SPH implementations utilise an
artificial viscosity to capture shocks \citep[but see
also][]{inu02}. This artificial viscosity can introduce numerical
momentum and angular momentum transport, and spurious energy
dissipation.  Indeed, \citet{sd01} found that the angular momentum of
a simulated galaxy increased when they used higher resolution.
Another problem arises from SPH's intrinsic smoothing
properties. Since SPH represents a fluid element by smoothing over
neighbouring particles, it is not well suited for treating large
density and velocity gradients.  This can be a serious problem when a
cold gas disc forms through radiative cooling in an ambient hot gas at
the virial temperature.  In this situation, the cold gas disc is much
denser than the hot gas and generally rotates faster than the ambient
hot medium.  In addition, because star formation can lead to a
decrease in the number of particles in the disc ($\sim$ 90 \% of
baryonic matter becomes stars in a disc), the effective spatial
resolution degrades with time, an effect which may play an important
role at low redshift.

In this paper we investigate angular momentum transfer from a cold gas
disc to the hot halo gas and its effect on the simulation outcomes.
Although alternative techniques based on Eulerian approaches coupled
with a grid refinement scheme --- Adaptive Mesh Refinement (AMR) ---
have been recently implemented in Cosmology
\citep[][and references therein]{abn00,tey02}, we concentrate here on 
numerical effects in SPH simulations of galaxy formation because SPH
is by far the most widely used method in this area.

The outline of this paper is as follows. A brief description of our
simulation code is given in Section 2. In Section 3, we carry out   
cosmological simulations of disc formation in a virialised halo,
and then demonstrate there is angular momentum transfer from the cold gas
disc to the ambient hot gas. A forensic study is performed in 
Section 4 using simplified simulations to find the source of the
problem and the dependence on the numerical resolution. 
In Section 5 we propose that decoupling of cold gas from ambient hot gas
can avoid the problems found in earlier sections. 
The effect of the decoupling is shown using
simplified simulations and cosmological simulations. 
The results are summarised and discussed in Section 6.

\section{THE CODE}

We use a modified and extended version of the parallel TreeSPH code
GADGET \citep{syw01}. Unless otherwise specified, we will use the novel
formulation of SPH that manifestly conserves energy and entropy when
appropriate \citep{sh02} throughout this paper.  
As shown in \citet{sh02}, this formulation greatly reduces numerical
inaccuracies compared to the more commonly used formulations.

GADGET employs an artificial viscosity which is the shear-reduced
version \citep{bal95, ste96} of the ``standard'' \citet{mg83} artificial
viscosity. 
\citet{lom99} and \citet{tha00} endorsed this form of the
artificial viscosity.  
We set the parameter $\alpha$ that appears in Eq. 27 of \citet{syw01}
to 0.75. We have checked that the choice of the value of this parameter 
hardly affects our results.

We also modify the gradient of the smoothing kernel from the public
version of GADGET according to \citet{tc92} to overcome the clumping 
instability \citep{ss81}. GADGET adopts the kernel, $W$, so-called
$B_2$-spline \citep{mon85}
\begin{equation}
W(r,h)=\frac{8}{\pi h^3}\left\{
	\begin{array}{ll}
		1 - 6 u^2 + 6 u^3 & \quad\mbox{for $0 \le u \le \frac{1}{2}$},\\
		2 (1 - u)^3       & \quad\mbox{for $\frac{1}{2} < u \le 1$},\\
		0                 & \quad\mbox{for $u > 1$},
	\end{array}
	\right.
\end{equation}
where $r$ and $h$ are the particle separation and smoothing length,
respectively, and $u=r/h$. Note that the smoothing kernel is defined
over the interval $[0,h]$ and not $[0,2h]$ as is more common.  The
kernel gradient vanishes at $u = 0$, i.e. the pressure gradient forces
between two close SPH particles vanishes in the limit of small
separation.  We modify the gradient for $u \le \frac{1}{3}$ so that
even close pairs of SPH particles continue to repel each other, but leave the
kernel itself unchanged,
\begin{equation}
\frac{\mbox{d$W$}}{\mbox{d$u$}} 
	= \frac{\mbox{d$W$}}{\mbox{d$u$}}\left(u=\frac{1}{3}\right) 
	= - \frac{16}{\pi h^3}, \ u \le \frac{1}{3}.
\end{equation}

Except in cosmological simulations, where we solve for the ionisation, 
we assume a fixed mean molecular weight, $\mu = 0.59$, corresponding
to a fully ionised gas of primordial composition.  We
use the adiabatic index, $\gamma = 5/3$ throughout this paper.

\section{COSMOLOGICAL SIMULATIONS OF DISC FORMATION} \label{cos1}

In order to study disc formation and to concentrate on the
investigation of numerical angular momentum transfer, we wish to avoid
the physical angular momentum losses which occur during hierarchical clustering as
much as possible while retaining a degree of realism.  To meet these requirements,
we choose a halo,  from a pre-existing $N$-body simulation, that
is known to have a quiet merger history -- the redshift of the last
major merger is larger than 1. In addition the gas is not
allowed to cool radiatively until after $z =1$. After this both cooling and
star formation are allowed. This procedure completely suppresses the
early formation of proto-galactic clouds and leads to quiescent gas accretion
for $z < 1$, with the result that a disc forms with a reasonable size
\citep{wee98, eew00}.  The details of the simulation are described in
the following subsections.

\subsection{Initial conditions}

The background cosmology that we assume is a low-density flat CDM 
universe ($\Lambda$CDM). This model is currently the favourite
amongst hierarchical clustering models. We use the following choice of
the cosmological parameters: $\Omega_0 = 0.3, h \equiv H_0/100$ km
s$^{-1}$ Mpc$^{-1}$ = 0.7, $\lambda_0 \equiv \Lambda_0/(3 H_0^2) = 0.7$,
and $\sigma_8 = 0.9$. 
The baryon density, $\Omega_{\rm b}$ is set to 0.04 \citep{net02}.

To generate our initial conditions, we use the resimulation technique
introduced by \citet{fews96}.  We first perform a dark matter only
simulation in a $35.325 h^{-1}$ Mpc periodic cube. On this scale, the
density fluctuations are still in the linear regime at $z = 0$. Having
completed this simulation, we then select a dark halo that has a
quiet merger history. The halo's mass is about $1.3 \times 10^{12}
h^{-1} M_{\odot}$ within the sphere having virial overdensity,
$\delta_{\rm vir} = 337$ at $z = 0$.  To make the new initial
conditions, the initial density field of the parent simulation is 
recreated and appropriate additional short wavelength perturbations
are added to the region out of which the halo forms.  In this region
we also place SPH particles in a ratio of 1:1 with dark matter
particles.  The region external to this was populated with high mass
dark matter particles whose function is to reproduce the appropriate
tidal fields.  The initial redshift of the simulation is 50.  The
masses of the SPH and high resolution dark matter particles are $\sim
2.6 \times 10^6 h^{-1} M_{\odot}$ and $\sim 1.7 \times 10^7 h^{-1}
M_{\odot}$, respectively.

The gravitational softening length for the SPH particles is kept fixed
in comoving coordinates for $z>3$ and after this, it is frozen in
physical units to a value of 0.5 kpc, in terms of the `equivalent'
Plummer softening given in \citet{syw01}. The gravitational force
obeys the exact $r^{-2}$ law at $r > 2.8 \epsilon$.  The softening
lengths for all particles are defined as $\epsilon_{\rm p} =
\epsilon_{\rm sph}
\times (m_{\rm p}/m_{\rm sph})^\frac{1}{3}$, where $m_{\rm p}$ is the
particle mass of a particle `p' and $\epsilon_{\rm sph}$ and $m_{\rm
sph}$ are the softening and mass of the SPH particles.  We do not
allow the smoothing length to become smaller than a minimum value 
of  $h_{\rm min} = 1.4 \epsilon$.

\subsection{Cooling and star formation}

For $z < 1$ the cooling/heating rate and ionisation state of each
particle are calculated assuming collisional ionisation equilibrium and
the presence of an evolving but uniform UV background \citep{hm96} by
using the fitting formula provided by \citet{the98}.  A primordial
composition for the gas is assumed.  Inverse Compton cooling is
also considered at $z < 1$ although the effect is minor.  Since we do
not include molecular cooling, the coolest gas typically has a
temperature $T_{\rm cold} \simeq 10^4$~K. We define the gas in an
overdense region that has a temperature $T < 3 \times 10^4$~K as
``cold gas.''

Cold gas particles are eligible to form stars when the following
criteria are satisfied: (i) the gas particle is in a converging flow
($\nabla \cdot \mbox{\boldmath $v$}_i < 0$) and (ii) the density of
the gas particle is above a threshold density ($\rho_i > \rho_{\rm
th}$).  We use $\rho_{\rm th} = 5 \times 10^{-25}$ g cm$^{-3}$.  The
value that we adopt is higher than the typical value used in other
cosmological simulations, $2 \times 10^{-25}$ g cm$^{-3}$
\citep{kwh96}. This choice allows us to have sufficient cold gas to
observe the numerical effect on the cold phase.  Note that
\citet{buo00} found that the criterion on the velocity divergence has
no sizeable effect on their results, and so criterion (i) may not
be needed.  We ignore the Jeans condition, used by some other authors
\citep[e.g. ][]{kwh96}, for reasons detailed below. 

The Jeans condition is usually denoted as
\begin{equation}
h_i/c_{\rm s} > t_{\rm dyn}, 
\label{jeans1}
\end{equation}
where $h_i$, $c_{\rm s}$, and 
$t_{\rm dyn} = (4 \pi G \rho_i)^{-\frac{1}{2}}$ are the smoothing
length, the sound speed, and the dynamical time of the gas particle
`$i$', respectively. Since the $h_i$ have no direct physical significance
in the SPH formalism, and depend for example on the particle mass, 
adopting such a Jeans condition, as given above, would  introduce an 
unphysical resolution dependence into the simulations.

In fact this Jeans condition should be regarded as determining
the resolution limit rather than as a star formation criterion.
\citet{bb97} have shown that if the minimum resolvable mass $\sim 2
N_{\rm ngb} m_{\rm sph}$, where $N_{\rm ngb}$ is the number of
neighbours used in the SPH calculation, becomes larger than the local
Jeans mass, $\sim G^{-\frac{3}{2}}\rho^{-\frac{1}{2}}c_{\rm s}^3$,
artificial fragmentation may occur, and real fragmentation will
definitely be suppressed.  In our adopted SPH implementation, the
smoothing length, $h_i$, is defined as $(4 \pi/3) {h_i}^3 \rho_i =
m_{\rm sph} N_{\rm ngb}$, and one finds combining these relations
that the above resolution limit is equivalent to
\begin{equation}
\frac{h_i}{c_{\rm s}} < \pi^{\frac{1}{2}}\left(\frac{3}{\pi}\right)^{\frac{1}{3}} t_{\rm dyn}
\simeq \pi^\frac{1}{2} t_{\rm dyn}. 
\label{jeans2}
\end{equation}
It is clear that the condition represented by Eq. \ref{jeans1} is
hardly satisfied if the condition represented by Eq. \ref{jeans2} is
satisfied.  We adopt $N_{\rm ngb}$ = 40 in this paper, and this gives,
for a gas temperature, $T = 10^4$ K, that for densities below
$\rho_{\rm max} \sim 1.2 \times 10^{-25}$g cm$^{-3}$  the SPH treatment of the gas
is reliable.  Note that this is well below our threshold density
$\rho_{\rm th}$ for star formation. Therefore, our results may be
affected by artificial effects because of insufficient mass resolution.

When a gas particle is eligible to form stars, the star formation rate
(SFR) for a particle `$i$' is
\begin{equation}
\frac{{\rm d}\rho_*}{{\rm d}t} = c_* \frac{\rho_i}{t_{\rm dyn}}, 
\label{sf}
\end{equation}
where $c_*$ is a dimensionless SFR parameter. This formula corresponds
to the Schmidt law that implies an SFR proportional to
${\rho_i}^{1.5}$. The value of $c_*$ controls the star formation
efficiency.  The physics of star formation is not understood well
enough to predict the value of this parameter. It is known for spiral
galaxies that the star formation time-scale is long compared to the
dynamical time-scale, so to mimic this, the value of $c_*$ must be
significantly less than unity. We assume a value of $c_* = 0.05$
throughout this paper. This is consistent with the relatively small
values of $c_*$ used in simulations of disc formation
\citep[e.g.][]{kat92,wee98,tc00}.  For an SPH particle, of mass
$m_{\rm gas}$, which is eligible to form a star, of mass $m_*$, the
probability of this event occurring during a time-step $\Delta t$ is
given by
\begin{equation}
p_{\rm sf} = {m_{\rm gas}\over m_*}\Bigg[1-\exp\left(-\frac{c_*\Delta t}
{t_{\rm dyn}}\right)\Bigg].
\label{prob}
\end{equation}

We use two star formation recipes to study the dependence of the
results on the star formation scheme and also the number of particles
left in a cold gas disc.  In the first recipe, a gas particle is
completely converted into a stellar particle during $\Delta t$ so that
$m_{\rm gas}/m_*=1$ in Eq. \ref{prob}.  We call a simulation
using this star formation scheme a ``conversion run''.  The other
recipe allows an initial SPH particle to spawn up to three stellar
particles with mass of $m_{\rm sph}/3$, where $m_{\rm sph}$ is the
original mass of the SPH particle, so that the values possible for
$m_{\rm gas}/m_*$ are 3, 2 or 1. This scheme reduces the rapid
decrease in the number of cold gas particles in the disc due to star
formation, and helps counter the large drop in the SPH spatial
resolution which otherwise occurs as the cold gas is used up. The
simulation which employs this scheme is dubbed a ``spawning run.''

\subsection{Results}

\begin{figure}
\includegraphics[width=8cm]{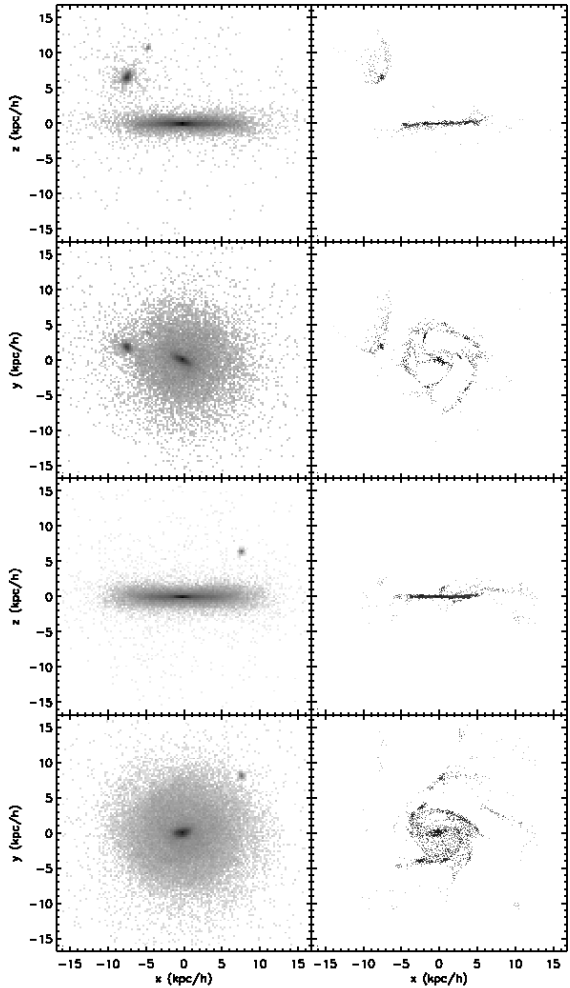}
\caption{Distribution of stars and cold gas in the galaxy at $z = 0$. 
	The upper 4 panels and the lower 4 panels are for the
	conversion run and the spawning run respectively. The left
	panels show the stellar distributions and the right panels
	show the cold gas distributions.  The greyscale is coded by
	surface density and three dimensional density for stars and
	cold gas, respectively. The $z$ direction is chosen so that
	the $z$-axis becomes parallel to the stellar angular momentum
	of each galaxy. }
\label{z0}
\end{figure}
\begin{figure}
\includegraphics[width=8cm]{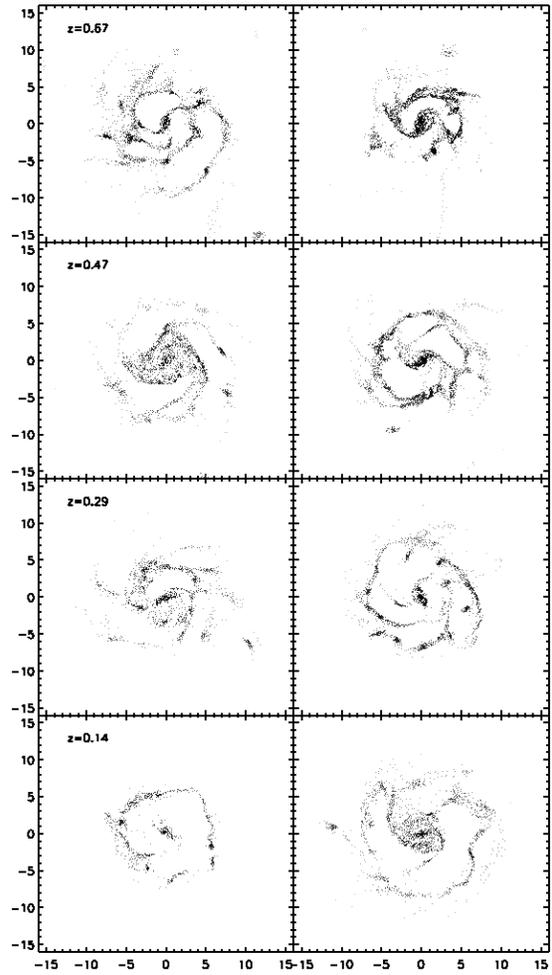}
\caption{Face on views of the distribution of the cold gas. 
	The left and right panels show the galaxy in the conversion run and 
        spawning run respectively.}
\label{redshift}
\end{figure}
In Fig. \ref{z0}, we show the distributions of stars and cold gas in the
conversion and spawning runs at $z = 0$. The galaxy in the spawning run
has a smoother stellar distribution than in the conversion run as
expected. Although the galaxy in the conversion run has a stronger bar,
we cannot find any significant difference in the distribution of the
stars when we compare the surface density profiles of these galaxies.  

However, the morphologies of the cold gas discs are quite
different. The cold gas disc in the conversion run has a core and ring
structure: the cold gas disc has large holes and most of the cold gas
particles are found in dense filaments.  The cold gas disc in the
spawning run is perhaps more realistic with spiral arm-like features,
though a large fraction of the cold gas lives in the arms. One might
think that the fragmentation is caused by the Toomre
instability. However, we find that when we calculate the value of
Toomre's $Q$-parameter for the azimuthally averaged surface gas
density, that $Q > 1$ is satisfied everywhere and at all
redshifts. Because of efficient star formation, the gas surface
density never reaches high values.  Having ruled out the Toomre
instability we need to examine the time evolution of the gas
distributions to understand the physical or numerical mechanisms that
cause the break up of the cold gas discs.

The distributions of the cold gas particles are shown in
Fig. \ref{redshift} for several redshifts in each simulation.  We find
that the core--ring structure is very common, regardless of the star
formation scheme. The spawning galaxy has a larger-sized cold gas disc
at lower redshift than the conversion galaxy. This is consistent with
the idea that there is angular momentum transfer away from cold gas in
a way which depends on the number of the cold gas particles.

\begin{figure*}
\includegraphics[width=16cm]{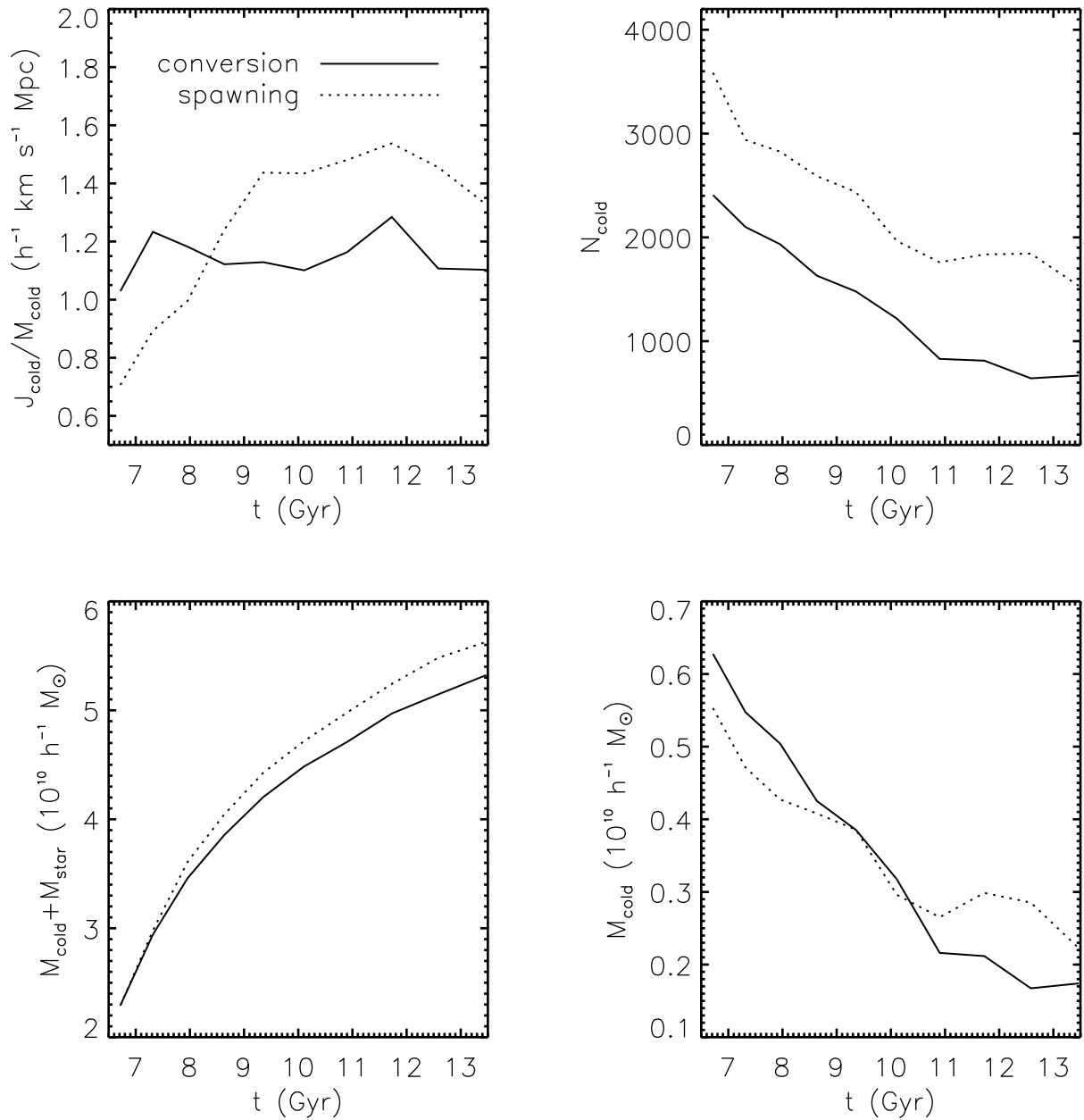}
\caption{Evolution of the cold gas disc. The upper left, upper right, lower
 right, and lower left  panels show the specific angular momentum of the
 cold gas disc, the number of cold gas particles in the discs, the
 mass in the stars and cold gas (integrated cooling rate), and the mass
 of the cold gas disc, respectively, as a function of the age of the
 universe. The solid and dotted lines indicate the conversion and
 spawning run, respectively.  
}
\label{jcold1}
\end{figure*}
\begin{figure}
\includegraphics[width=8cm]{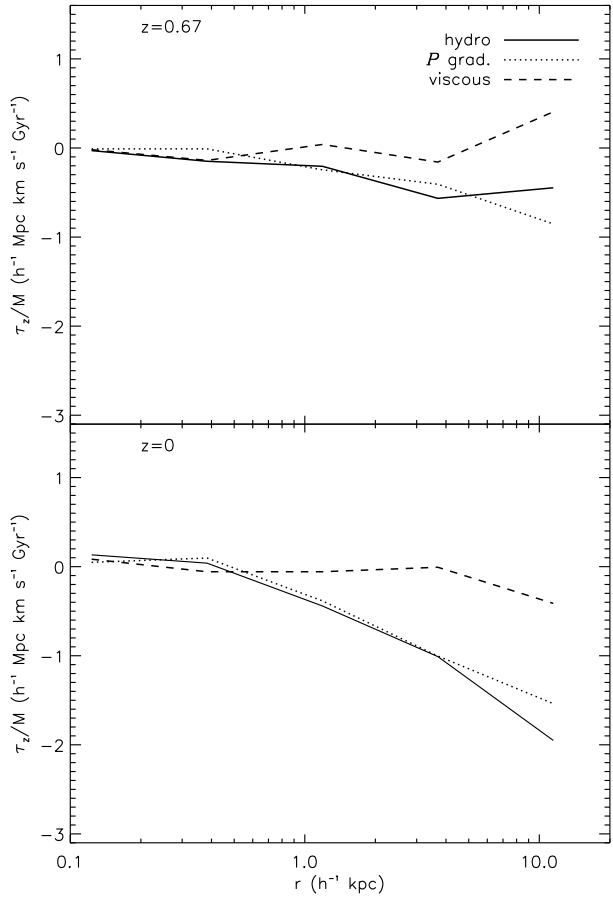}
\caption{The hydrodynamic torques acting on the cold gas particles in the
 disc. The results for the spawning run are shown. 
 Azimuthally averaged specific torques parallel to the angular
 momentum of the disc are plotted as functions of radius. The solid line
 represents the total hydrodynamic torque. This torque is decomposed into
 the torque from pressure gradient force (dotted) and that from the
 artificial viscosity (dashed). }
\label{torque1}
\end{figure}

Since star formation turns low angular momentum gas particles into
collisionless stellar particles and the gas particles that accrete
onto the disc later tend to have larger angular momenta, the specific
angular momentum of the cold gas disc should monotonically increase
with time.  We plot the evolution of the specific angular momentum of
the cold gas disc in Fig. \ref{jcold1}. Here, we consider the material
in a sphere of radius 20 $h^{-1}$ kpc, which is centred on the galaxy
centre at each redshift.  The specific angular momentum for the
conversion run shows surprisingly little evolution, being nearly
constant. In contrast, the angular momentum of the cold gas disc in
the spawning run increases monotonically except for the last few
gigayears. This indicates that more angular momentum is lost from the
cold gas disc in the conversion run compared to the spawning run.

The lower left panel of Fig. \ref{jcold1} shows the integrated cooling
rates (the mass in stars and cold gas at each redshift) in the
galaxies.  The figure shows that the cooling in the conversion run is
suppressed relative to the spawning run.  This indicates that a
significant amount of angular momentum from the cold gas disc in the
conversion run has been transfered to the ambient hot gas, and that
the cooling rate has decreased because the hot gas has been puffed up
by this ``angular momentum feedback.''  Since the difference in the
amount of cold gas in the discs between the two simulations is small,
the difference in the masses of the cooled baryon (i.e. cold gas and
stars) is mainly due to the difference in the masses of stellar
discs. By comparing the evolution of the angular momenta of the cold
gas discs and the number of cold gas particles in the discs (upper
right panel of Fig. ~\ref{jcold1}), one might draw a naive conclusion
that at least 2000 cold gas particles are needed to suppress the
angular momentum transfer.  However, we have to understand the
mechanism that causes the angular momentum transfer and the
fragmentation of the cold gas disc before we can reach definite
conclusions.

To this end, we next investigate the hydrodynamic torques acting on
the cold gas particles. In Fig. \ref{torque1}, we plot the azimuthally
averaged specific hydrodynamic torques acting on the cold gas
particles for the spawning simulation for 5 radial bins.  The negative
value indicates that the torque spins down the rotation of the disc.
As expected from Fig. \ref{jcold1}, the absolute magnitude of the
torque is larger at lower redshift. Surprisingly, the torque is
dominated not by the artificial viscosity but by the contribution to
the force due to pressure gradients. The total hydrodynamic torques
normalised by the angular momenta of the cold gas discs ($\tau_z/J_z$)
at $z = 0$ are -0.84 and -0.90 Gyr$^{-1}$ for the conversion and
spawning runs, respectively.  This means that the hydrodynamic torque
can stop the rotation completely in only $\sim 1$ Gyr.  While the
number of the cold gas particles in the spawning run is more than
twice as large as that in the conversion run, both discs receive
comparable torques at $z = 0$.  This implies that the strength of the
hydrodynamic torque is defined by the distribution (morphology) of the
cold gas as well as the number of the cold gas particles in the disc.

\begin{figure}
\includegraphics[width=8cm]{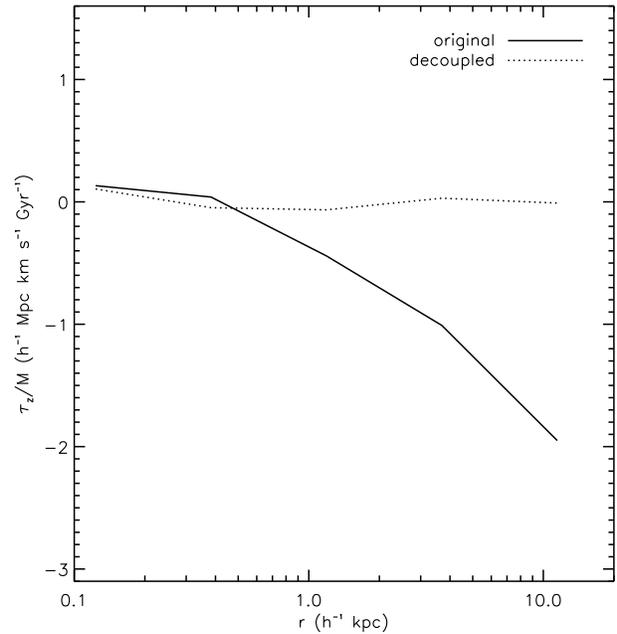}
\caption{The hydrodynamic torque acting on the cold gas particles in the
 disc in the spawning run.  The solid line is identical to the solid
 line in the lower panel of Fig. \ref{torque1}. The dotted line shows the
 torque when we ignore the interaction between the cold gas 
($T < 3 \times 10^4$ K) and the hot gas ($T > 10^{5}$ K) particles. }
\label{torque2}
\end{figure}

To know whether there is significant angular momentum transfer between
the cold gas particles at different radii, we calculate the
instantaneous hydrodynamic torque acting on the cold gas disc ignoring
the interaction between the cold ($T < 3 \times 10^4$ K) and hot ($T >
10^5$ K) phases. Fig. \ref{torque2} shows the original torque that is
identical to the solid line in the lower panel of Fig. \ref{torque1}
and the torque ignoring the hot gas. We find that the hydrodynamic
torque becomes almost 0 at all radii when we decouple instantaneously
the cold and hot phases. There is no significant transport of angular
momentum within the cold gas itself.  This confirms results from
previous studies, which insist that the angular momentum transfer in
the galactic disc itself cannot be a serious problem over a Hubble
time when the shear-reduced artificial viscosity is adopted
\citep[e.g.][]{ste96}.  Unfortunately, all test simulations have been
performed in the absence of a surrounding (hot) medium.

Since our results indicate that the artificial viscosity is not very
important for the angular momentum transfer, the transfer may be
caused by fragmentation or may cause the fragmentation.  We will
investigate this point using simplified simulations in the next
Section. We note that, the gravitational torque acting on the cold gas
disc is also significant when the cold gas is fragmented.

\section{SIMPLIFIED SIMULATIONS}

\begin{figure*}
\includegraphics[width=18cm]{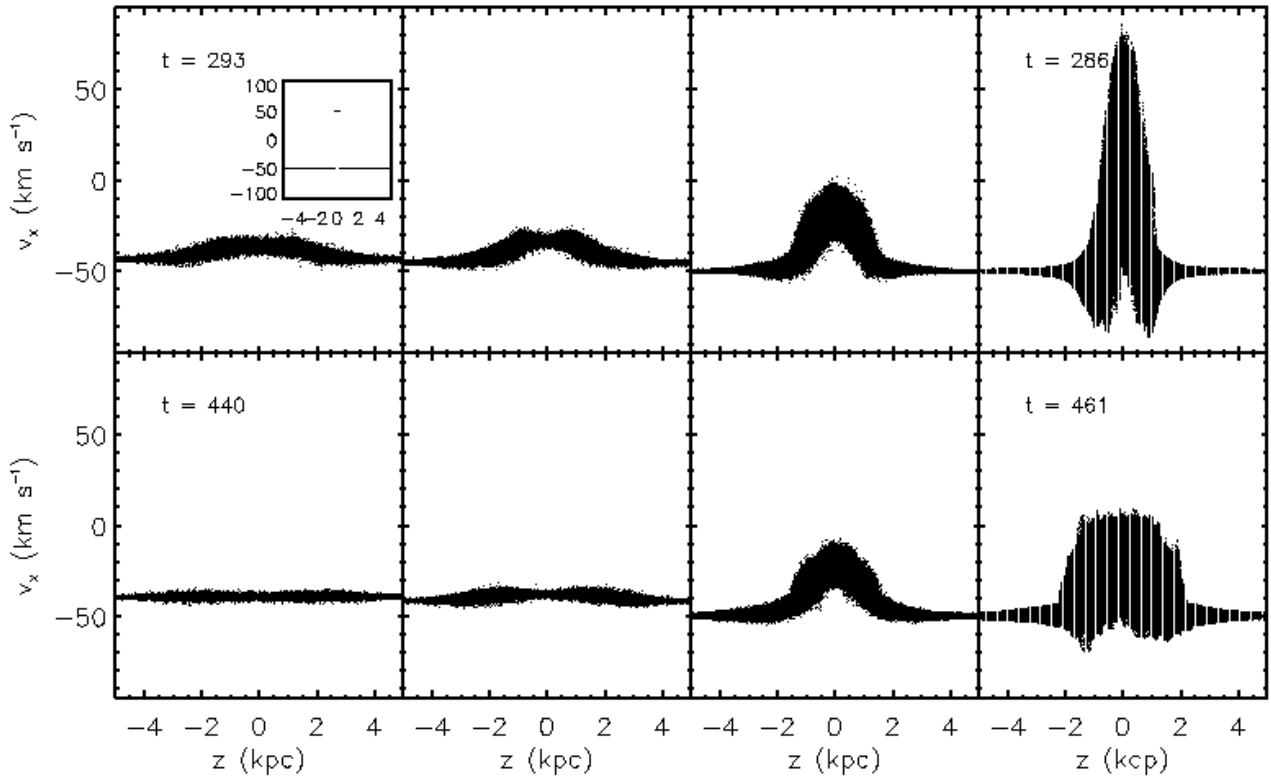}
\caption{
The evolution of the velocity profile in the single temperature shear tests. 
The results from the SPH runs are shown in the three left-hand columns
($(N_{\rm gas},N_{\rm ngb})=(32^3,40), (64^3,40)$ and $(64^3,320)$
from left to right) and the results from the FD code are shown in the
right panels. The top and bottom panels corresponds the outputs at 
$t = 293,$ and 440 Myr for the SPH simulation and 
$t = 286,$ and 461 Myr for the FD simulation, respectively, and an insert
in the top left panel shows the initial velocity field.
}
\label{singlev}
\end{figure*}

\subsection{Shear flows}

Several authors have presented shear flow tests using SPH, and their
results are promising \citep{lom99, tha00}. However, these studies
have focused primarily on the variation resulting from using different
SPH implementations. Here, we consider the impact of varying numerical
resolution upon the momentum transferred across sharp velocity
gradients, with or without associated sharp density gradients.

\subsubsection{A single temperature test}

We use simulations of a periodic cube of side 10 kpc, containing
$10^{10} M_{\odot}$ of gas, to investigate the SPH transfer of 
momentum across a discontinuity in the velocity field. The gas is all
given a temperature of $10^6$ K. Particles in the central slab with
$|z|<0.3$ kpc are given a velocity of $v_x=50$ km s$^{-1}$, and the
remaining gas is set up with $v_x=-50$ km s$^{-1}$. 
At this relative velocity of 100 km s$^{-1}$, it takes $\sim 100$ Myr
to cross the box. The self-gravity of the gas is ignored for
simplicity, and replaced with an external potential of the form
\begin{equation}
\Psi(z) = -10000\left[\cos\left(\frac{2 \pi z}{L_{\rm box}}\right)-1\right]
 ({\rm km} \ {\rm s}^{-1})^2, 
\label{extpot}
\end{equation}
where $L_{\rm box}$ is the side length of the simulation box.
The choice of external potential is not very important because the transfer
of momentum is not greatly affected by the size of the
instabilities that it suppresses.
To generate relaxed initial conditions, we first distribute particles
using the rejection method assuming hydrostatic equilibrium to
calculate the density.
This system is evolved without any shear, while damping the particle
temperature and
velocity until a relaxed state is reached. Then the shear is introduced
and the tests are commenced.
Simulations have been run with $(N_{\rm gas},N_{\rm ngb})=(32^3,40),
(64^3,40)$ and $(64^3,320)$.
A single variant of SPH has been used for the runs in this subsection,
but we do consider some other popular flavours with different
symmetrisations in subsection~\ref{sssec:othersph}.

We have also used the three-dimensional
Eulerian fixed-grid hydrodynamic code described by \citet{qis96} to
provide a comparison. This finite difference (FD)
code employs a Riemann solver to compute the
numerical viscosity, thus removing the need for an artificial
viscosity, and has already been used to simulate gas
stripping from a galaxy by the ram pressure of the intracluster medium (ICM)
\citep{qmb00} and the evolution of a bubble in the ICM \citep{qbb01}.
We set up the same initial conditions for the shear test on $151^3$ cells.
This number provides ample resolution and ensures that each cell hosts
only one phase initially.  
Random noise is added to the density in each cell ($\Delta
\rho/\rho \le 0.01$), otherwise nothing will happen. 

The evolution of the velocity profile for each simulation is shown in
Fig.~\ref{singlev}. For the SPH runs employing only 40 neighbours, the
velocity shear decays rapidly owing to momentum transfer across the
shear boundary. Using 320 neighbours, the transfer of momentum is
significantly suppressed.  In the FD simulation, the width of the
distribution of $v_x$ values at a particular $z$ coordinate reflects
large scale turbulence that is not apparent in the SPH runs. The peak
velocities are typically larger and the velocities of the layers
distant from the contact surfaces remain intact even after 5 box
crossings of evolution.

\begin{figure}
\includegraphics[width=7.5cm]{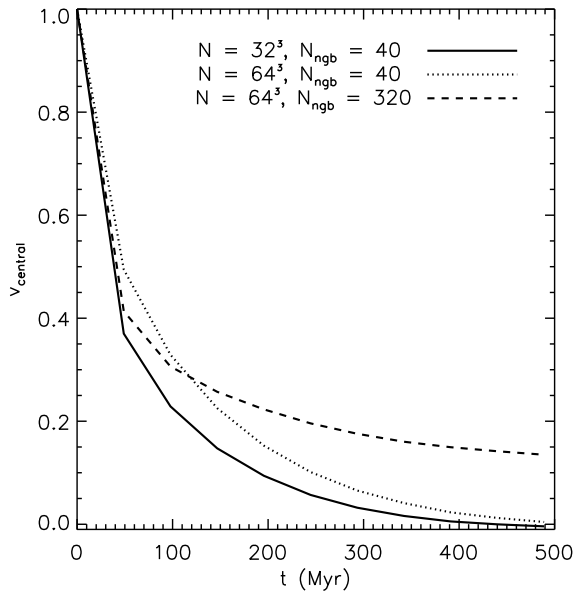}
\caption{
Evolution of the mean $x$ velocity of the cold phase in units where 1
corresponds to the initial velocity and 0 is reached when all shear
has gone.  Solid, dotted, and dashed lines indicate the shear
simulations using $(N_{\rm gas},N_{\rm ngb})=(32^3,40), (64^3,40)$ and
$(64^3,320)$, respectively. At early times, the momentum loss is
caused largely by artificial viscosity. As the relative velocity
decreases, pressure gradients provide the more important deceleration.
}
\label{killer}
\end{figure}

By its very nature, SPH is not well-suited to solving problems
involving large discontinuities. In the following subsection we will
show more vividly how the SPH and FD methods give qualitatively
different results regarding turbulence. However, it is instructive to
understand the difference between the SPH runs considered above. As we
have seen, the hydrodynamical force can be split into two components:
pressure gradients and artificial viscosity. The latter depends upon
the relative velocity of the fluids and the size of the interacting
volume. This boundary layer should be the same for the $(32^3,40)$ and
$(64^3,320)$ simulations, because $N_{\rm ngb}/N_{\rm gas}$ is
identical. However, the first and third columns in Fig.~\ref{singlev}
provide drastically different results, so we can infer that the
pressure gradients must be behind this variation. Fig.~\ref{killer}
shows at early times that these two simulations do lose momentum at a
similar rate. The $(64^3,40)$ run, and its smaller boundary layer,
initially slow down less rapidly.  As the relative velocity of the
gases decreases, the dominant force leading to deceleration becomes
that coming from pressure gradients. This depends on $N_{\rm ngb}$,
such that more neighbours yield smaller decelerations. Thus, it seems
to be noisiness in the SPH smoothing of variables which gives rise to
these pressure gradients and a significant proportion of the
deceleration. The late time evolution in Fig.~\ref{killer} shows the
reduced deceleration of the $(64^3,320)$ run relative to the other
two. Note that increasing the number of neighbours in SPH calculation
significantly slows down the simulation as well as decreasing the mass
resolution.
\citet{ii02} pointed out that density errors in SPH simulations of shear
flows can be substantially suppressed by treating particle velocity
and fluid velocity separately.  However, their method is quite slow
and so far only works with a constant smoothing length (Imaeda,
private communication).

\subsubsection{A two phase gas}

Now that we have seen how SPH behaves when there is a sharp velocity
gradient, it is worth investigating the more realistic case of a cold
slab of gas moving relative to a hotter medium. To this end, we have
recreated initial conditions with the central slab of gas having
$T_{\rm cold} = 10^5$ K and the remaining gas left at $T_{\rm hot} =
10^6$~K.  The sound crossing time for the cold slab (i.e. $0.6$
kpc/$c_{\rm s}$) is $\sim 12$ Myr, and $\sim$ 57 \% of the mass is in
the cold phase.  This time, when creating the initial conditions, only
the temperature of the cold phase and velocities were damped, although
a maximum temperature of $T_{\rm hot}$ is also imposed.  Simulations
were performed using $16^3, 32^3$, and $64^3$ particles.

\begin{figure}
\includegraphics[width=8cm]{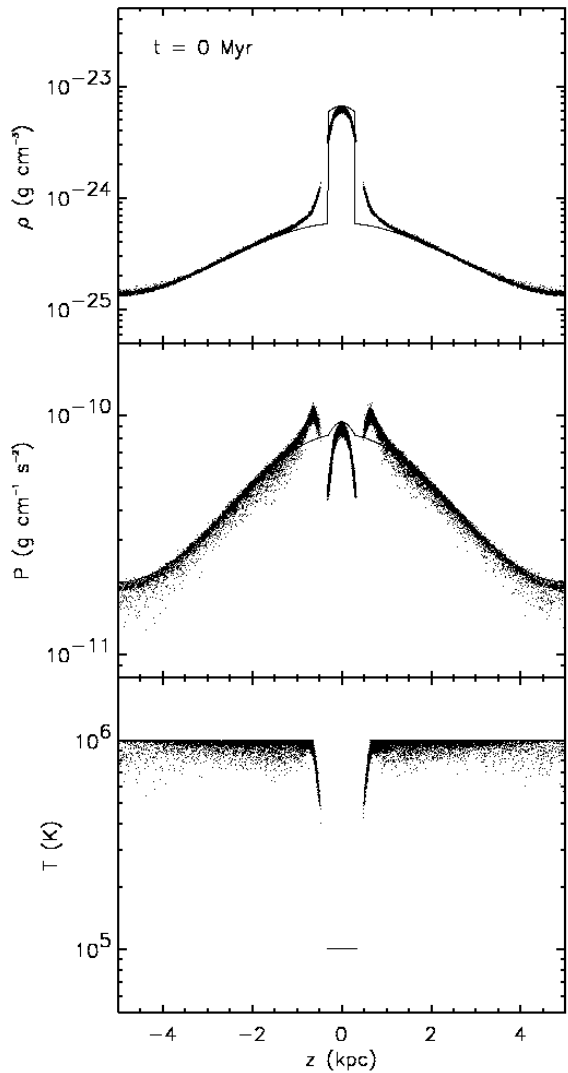}
\caption{
Densities (top), pressures (middle), and temperatures (bottom) of the
particles in the relaxed initial conditions with $N = 32^3$.  The
target density and pressure calculated from the external potential are
given by the solid lines.  }
\label{rptnoshear}
\end{figure}

\begin{figure}
\includegraphics[width=9cm]{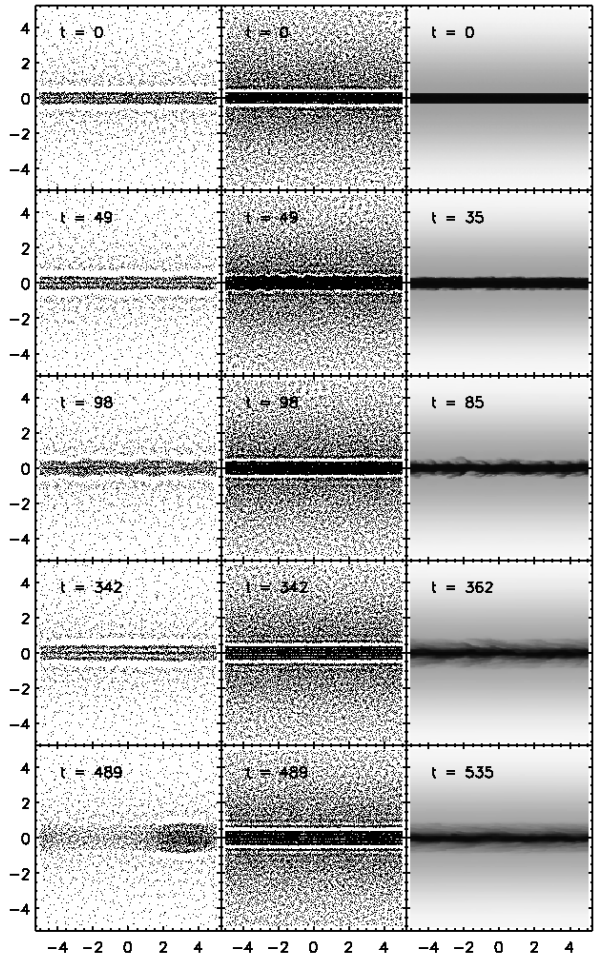}
\caption{
The evolution of the $x$--$z$ projections of particles in the shear 
simulations using $16^3$ SPH particles (left-hand column), $32^3$
particles (middle column) and gas density in the FD code (right column).
}
\label{sheargas}
\end{figure}

Fig.~\ref{rptnoshear} shows the densities, pressures, and temperatures
of the particles in the relaxed initial conditions using $N = 32^3$
particles.  It is apparent that the SPH density and pressure deviate
from the analytical curves near to the boundary. The presence of
features like these is inevitable with SPH \citep[see][]{pea99, rt01}.
Hot particles near to the cold dense slab overestimate their densities
and hence pressures. This causes the hot gas to expand away from the
dense region, adiabatically cooling in the process, and creating a gap
between the two phases which is clearly visible in
Fig.~\ref{sheargas}.  This figure also shows that the cold slabs are
divided into layers, the number of which depends upon the size of the
SPH smoothing length.

\begin{figure}
\includegraphics[width=8cm]{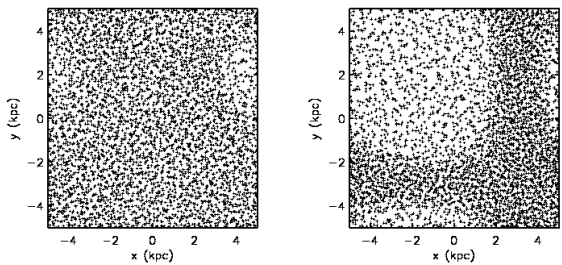}
\caption{
The $x$--$y$ projections of particles at $t = 342$ (left) and 489 Myr
 (right) in the shear simulation using $16^3$ particles.
}
\label{shear16gasxy}
\end{figure}
We now switch on the shear flow as before. The boundary layer is now
unstable to the Kelvin-Helmholtz instability \citep{ll87}.  The
presence of a fixed gravitational potential can in principle stabilise
long wavelength modes but for our configuration this effect can be
largely ignored.  The left-hand column in Fig.~\ref{sheargas} shows
the evolution of the $x$--$z$ projections of particles in the $N =
16^3$ simulation. It is apparent at later times that the cold slab is
breaking apart.  Instabilities are clearly visible at $t = 98$ Myr. At
$t = 342$ Myr an underdensity can be seen at ($x,y$) $\sim$
($4.5,2$). This subsequently grows and the cold phase at $t = 489$ Myr
no longer looks like a slab. This is confirmed in the face-on
projections shown in Fig.~\ref{shear16gasxy}.  Note that while the
morphology of the phases changes, the membership of each phase is
constant with cold particles remaining cold and hot ones hot. This
contrasts starkly with the behaviour of the FD simulation, shown in
the right-hand column of Fig.~\ref{sheargas}, where the turbulence
mixes the phases at the boundary between them.

In the middle column of Fig.~\ref{sheargas}, we show the particle
distributions in the $N = 32^3$ simulation. The instability appears at
$t = 49$ Myr, and then vanishes quickly. No significant evolution is
observed in either this run or that with $N = 64^3$, and in neither
case do the two phases mix at all.  Rerunning the $N = 32^3$
simulation with a cold gas slab of width 0.4 kpc, rather than 0.6 kpc,
does produce holes.  These runs lead us to conclude that the holes
seen here, and also in the simulations of disc formation, were formed
due to numerical artifacts. To avoid the formation of holes, the cold
phase must contain enough particles. This is a more important
consideration than the number of SPH neighbours being used.  It is
quite hard to maintain enough cold gas disc particles to prevent hole
formation in galaxy formation simulations, especially when star
formation is included.

\begin{figure}
\includegraphics[width=7.5cm]{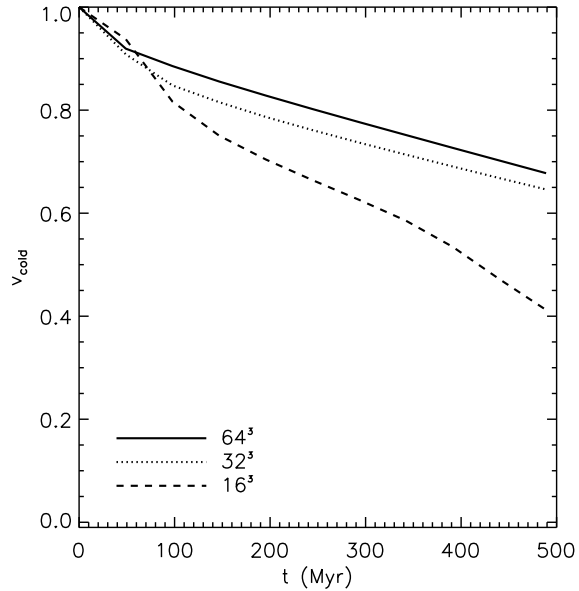}
\caption{
Evolution of the mean $x$ velocity of the cold phase in units where 1
corresponds to the initial velocity and 0 is reached when all shear
has gone.  Solid, dotted, and dashed lines indicate the shear
simulations using $64^3, 32^3$, and $16^3$ particles,
respectively. The inflection point at $t \approx 350$ Gyr for the
lowest resolution run results from hole formation in the cold slab and
ram pressure becoming important in enhancing the momentum loss.  }
\label{velocity1}
\end{figure}
One consequence of hole formation is shown in Fig.~\ref{velocity1},
where the evolution of the mean velocity of the cold phase in the $x$
direction is traced for each SPH simulation.  Since there is no mixing
of the two phases, the evolution of the mean velocity is equivalent to
the evolution of the momentum of the cold slab.  The momentum of all
the gas is well conserved, so any change in the velocity of the cold
phase should be regarded as the result of a momentum transfer with the
hot phase.  There is a resolution dependence of the size of the
initial deceleration when the slab is perturbed by the
Kelvin-Helmholtz instability.  We have confirmed that this
deceleration is caused by the pressure gradient force rather than the
artificial viscosity.  After the initial deceleration, the slabs lose
their momentum at an almost constant rate, where the acceleration from
pressure gradients and artificial viscosity are both important.  For
the lowest resolution run, an additional feature in the momentum
evolution can now be seen at $t \sim 350$ Myr.  This corresponds to
the epoch when the hole forms in this run and ram pressure from the
hot intra-hole gas leads to an extra deceleration of the cold
material. Such an effect is not evident in the simulations using
$32^3$ and $64^3$ particles. Their results nearly converge, apart from
the initial deceleration, although presumably not to a realistic
answer given that the SPH scheme suppresses any phase mixing.

\begin{table}
\caption{
Hydrodynamic acceleration parallel to the $x$ direction acting on the
cold phase in each SPH simulation at $t = 489$ Myr.  The first column
indicates the number of particles in the simulation.  The second,
third, fourth and last columns show the total acceleration, the
acceleration from the artificial viscosity, the acceleration from the
pressure gradients, and the acceleration from the ram pressure
respectively.  The acceleration is normalised by the velocity of the
cold phase.  The unit is Gyr$^{-1}$. }
\label{acc64}
\begin{tabular}{@{}lcccc}
\hline
  & $\left(\frac{a_x}{v_x}\right)_{\rm total}$ &
  $\left(\frac{a_x}{v_x}\right)_{\rm AV}$ &
  $\left(\frac{a_x}{v_x}\right)_{\nabla P}$ &
  $\left(\frac{a_x}{v_x}\right)_{\rm RP}$ \\
\hline
$16^3$ & -1.9 & -0.21 & -1.7 & -1.0 \\
$32^3$ & -0.55 & -0.35 & -0.21 & 0.035 \\
$64^3$ & -0.59 & -0.32 & -0.27 & -0.011\\
\hline
\end{tabular}
\end{table}
In Table~\ref{acc64} we show the hydrodynamic acceleration parallel to
the $x$ direction acting on the cold phase in each run at $t = 489$
Myr.  As expected, the cold slab in the $N=64^3$ run receives weaker
deceleration from the artificial viscosity than in the $N=32^3$ run.
Although the cold phase in the $N=16^3$ run receives the weakest
deceleration from the artificial viscosity, we ignore it because the
morphology of the cold phase in this run is no longer a slab.
Interestingly, the deceleration from pressure gradients is greater in
the $N=64^3$ run than in the $N=32^3$ run.
We also estimate the contribution of ram pressure to the pressure
gradient-based deceleration by calculating
\begin{equation}
a_{\rm RP} = \frac{1}{\rho}\frac{\partial \rho v_x^2}{\partial x} 
\end{equation}  
using an SPH gather formulation. In the $N=16^3$ simulation, half of
the deceleration comes from ram pressure, while the contribution of
the ram pressure is negligible in other simulations. It proves that
the artificial hole formation in the smallest simulation significantly
enhances the momentum transfer.

The slabs in $N=32^3$ and $64^3$ receive larger acceleration from the
artificial viscosity than the pressure gradients.  We have checked,
however, that the slabs lose their momentum more quickly only by
pressure gradients when we switch off the artificial viscosity,
because the lack of the artificial viscosity increases the numerical
diffusion and the noisiness in the particle distribution.

\subsubsection{Other SPH implementations}\label{sssec:othersph}

While the numerical difficulties caused by sharp boundaries will
inevitably impact adversely upon all standard implementations of SPH,
it is useful to be aware of the variation in momentum transfer rates
within this set of algorithms.  We now show the results for two other
SPH implementations.  The first one employs an arithmetically
symmetrised equation of motion and an asymmetric form of the energy
equation. This was dubbed `energy: asymmetric' in \citet{sh02},
produced the best results among the conventional SPH family and has
been widely used in galaxy formation \citep{evr88, rs91, nw93, sm93,
hk97, tha00}.  The second SPH variant we choose is one that employs
the geometric mean to symmetrise the equation of motion and the energy
equation.  This formula was proposed by \citet{hk89}, and was called
`energy: geometric' in \citet{sh02} where it produced the worst
results in their tests.  In accordance with \citet{sh02}, we refer to
our default implementation as `entropy: conservation.'  Note that we
regenerate initial conditions for each run using the appropriate SPH
variant to generate relaxed initial conditions.

\begin{figure}
\includegraphics[width=7.5cm]{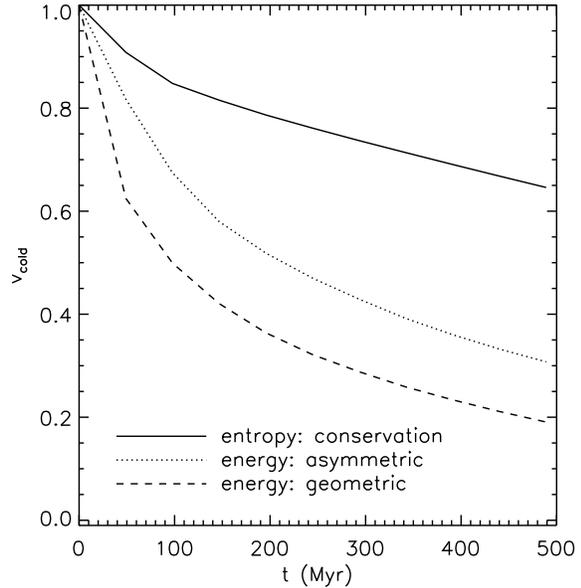}
\caption{
The same as Fig.~\ref{velocity1}, but for the different
implementations of SPH. In each simulation, $32^3$ particles are
used. The solid, dotted, and dot-dashed lines indicate ``entropy:
conservation'', ``energy: asymmetric'', and ``energy: geometric''
implementations, respectively. The solid line here is the same as the
dotted line in Fig.~\ref{velocity1}.  }
\label{velocity2}
\end{figure}

In Fig.~\ref{velocity2}, we plot the evolution of the mean velocity of
the cold slabs in the simulations adopting the above three flavours of
SPH.  The number of the particles in each simulation is $32^3$.  If we
assume that better algorithms lose less momentum, then we reach the
same conclusion as \citet{sh02} despite using a completely different
test, namely that `entropy: conservation' is the best and `energy:
geometric' is the worst.

\subsection{Disc formation in a rotating sphere} 

We now consider the idealised case of disc formation in a virialised
rotating sphere similar to that studied by \citet{nw93} and
\citet{tha00}. The purpose of this exercise is to obtain a
quantitative understanding of how the numerical effects studied in the
preceding subsections impact upon simulations that are closer to what
we expect occurs when galaxies form.

The initial conditions are created by placing particles on a cubic
grid with a spherical edge, and perturbing them radially to give a
density profile of the form $\rho(r) \propto r^{-1}$. Their velocities
are chosen so that the sphere will end up in solid-body rotation
around the $z$-axis; the initial angular momentum $J$ corresponds to a
value of the spin parameter, $\lambda=J|E|^{1/2}/(G M^{2.5}) \sim
0.1$, where $E$ is the total energy of the system.  The initial radius
of the sphere is taken to be 100 kpc, and its mass $M$ $10^{12}
M_{\odot}$, giving a free-fall time from the edge of the system of
about $524$ Myr.  A baryon fraction of 0.1 is assumed and equal
numbers of dark matter and gas particles are used.

In order to prevent the disc from becoming Toomre unstable, we impose
a high minimum temperature $T_{\rm min} = 10^5$ K.  This floor is
still well below the virial temperature of the system $T_{\rm vir}
\sim 3 \times 10^6$ K, and has the additional benefit of softening the
Jeans condition given in Eq. \ref{jeans2}.

The simulation is allowed to evolve for 1.25 Gyr without cooling to
let the hot gas reach equilibrium in the halo. Then radiative cooling
is switched on (using the cooling function computed for collisional
ionisation equilibrium by \citet{sd93} assuming a primordial mix of H
and He, and $\mu = 0.59$) and the system is followed until $t = 8$
Gyr.

\subsubsection{Cooling only simulations}

We first perform a series of cooling only (no star formation)
simulations using $N_{\rm gas} = 1736, 5544, 15408, 28624$, and 44442
SPH particles.  The same gravitational softening, $\epsilon = 2$ kpc,
is adopted for gas and dark matter particles for the simulation using
$2 \times 1736$ particles.  Softening lengths for other simulations
are chosen as $\epsilon = 2 \times (1736/N_{\rm gas})^\frac{1}{3}$
kpc.  The SPH smoothing length is allowed to decrease to $h_{\rm
min}=0$ in all runs because this choice results in better numerical
convergence.  Here, the relatively high temperature floor ($10^5$ K)
obviates the need to impose a minimum smoothing length.
\begin{figure}
\includegraphics[width=8cm]{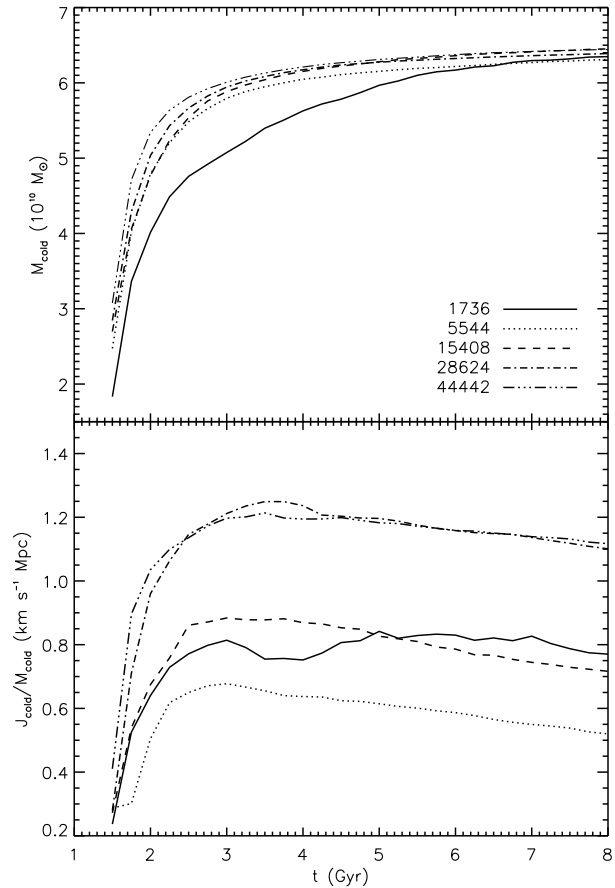}
\caption{
Evolution of the cold gas discs in the cooling only simulations. The
solid, dotted, dashed, dot-dashed, and triple-dot-dashed lines denote
simulations that employ $N_{\rm gas}$ = 1736, 5544, 15408, 28624, and
44442, respectively.  The upper panel shows the mass of the cold gas
disc as a function of time (integrated cooling rate).  The lower panel
shows the specific angular momentum of the cold gas disc.  }
\label{jcoldnw1}
\end{figure}

`Cold gas' is defined as gas that with a temperature lower than $1.3
\times 10^5$ K. The top panel of Fig. \ref{jcoldnw1} shows the
evolution of the cold gas disc mass. Higher resolution allows higher
central gas densities, so the better resolved runs have higher cooling
rates in the first $\sim 1$ Gyr. After this time, the simulations
using more than $\sim 15000$ gas particles create almost identical
disc masses, and that with 5544 SPH particles is only a few per cent
lower.

The evolution of the specific angular momenta of the cold gas discs,
shown in the lower panel of Fig. \ref{jcoldnw1}, is more varied. Since
the cooling is not calculated correctly in the simulation with 1736
gas particles, we will ignore this case.  Among the remaining
simulations, there is a monotonic increase of angular momentum with
increasing resolution and the evolution converges at $N_{\rm gas} =
28624$.  In all cases, the specific angular momenta are decreasing
functions after $t=3$ Gyr for all the simulations that resolve the
cooling adequately, despite the monotonic increase of specific angular
momentum with radius when the cooling is switched on.  As the total
angular momentum of the gas is well conserved, this implies that there
is some transfer within the gas.

In summary, for this particular test, at least 5000 total gas
particles are needed before the disc mass is well determined, while
more than 25000 are required to estimate the disc's angular momentum
reliably. Only about 60 per cent of these particles actually end up in
the disc itself.

\subsubsection{Simulations with star formation} \label{sftest}

\begin{figure*}
\includegraphics[width=16cm]{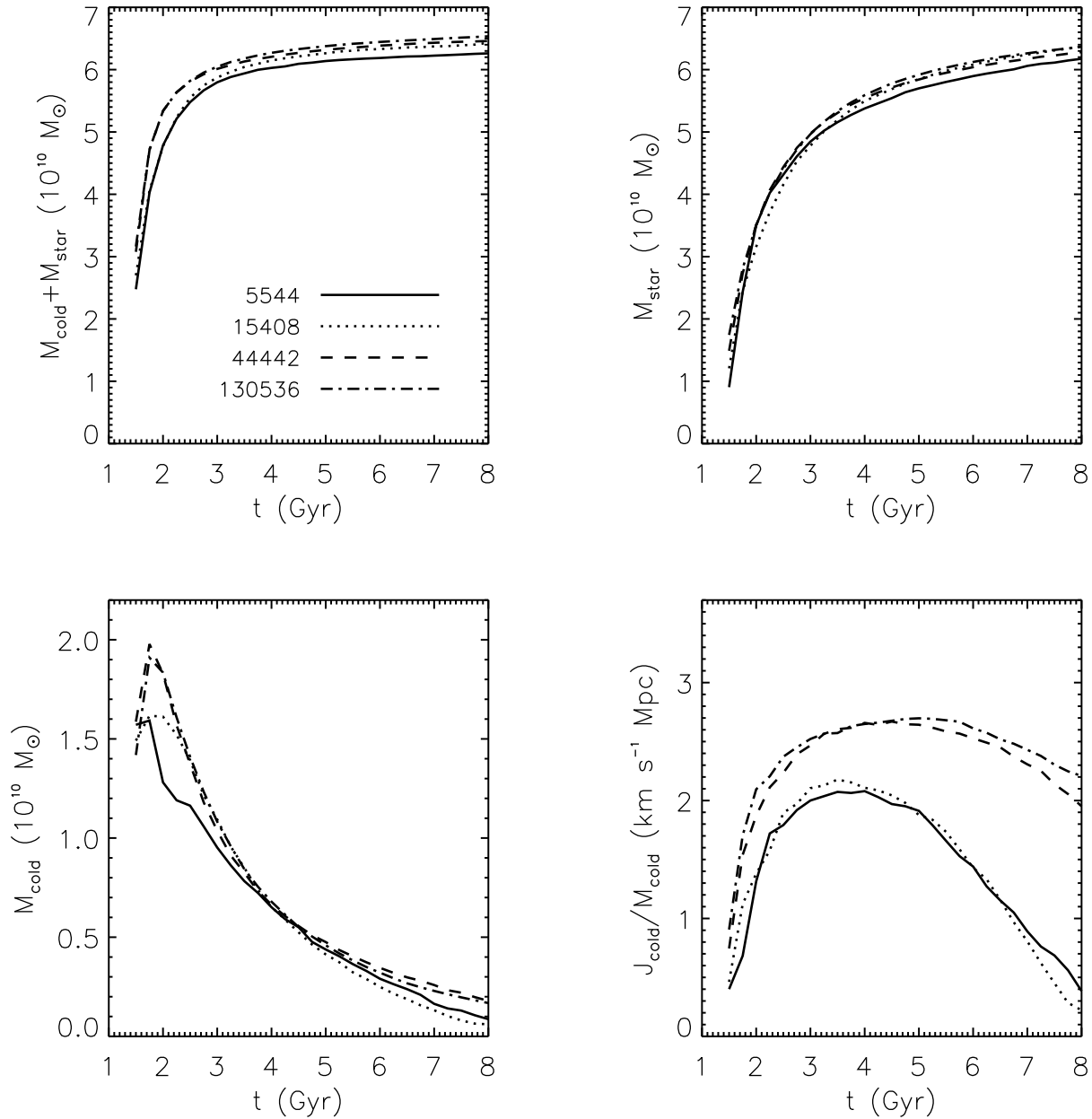}
\caption{
 Evolution of the cold gas discs in the simulation with star
 formation. The integrated cooling rate, the integrated star formation
 rate, the remaining cold gas mass, and the specific angular momentum
 of the remaining cold gas are plotted in the upper left, upper right,
 lower left, and lower right panels, respectively. The solid, dotted,
 dashed, and dot-dashed lines indicate the simulations with $N_{\rm
 gas} = 5544, 15408, 44442$, and 130536, respectively.  }
\label{jcoldstar}
\end{figure*}

As seen in Section~\ref{cos1}, the specific angular momentum of a cold
gas disc depends strongly on the number of particles it
contains. Thus, including star formation with an algorithm that
decreases the number of cold gas particles should exacerbate the loss
of angular momentum from the cold gas disc.  We now present
simulations similar to those of the previous section, but also
employing the `conversion' star formation scheme.  A lower threshold
density of $\rho_{\rm th} = 10^{-25}$ g cm$^{-3}$ was adopted, because
the cold gas disc is more diffuse than in the cosmological simulation
as a result of the higher minimum temperature allowed for the cold gas
($10^5$ K). We only show results from simulations for which a minimum
smoothing length was not imposed, and omit both the lowest resolution
simulation, in which cooling was not properly followed, and the
$N_{\rm gas} = 28624$ simulation, which is abased by star formation to
the low resolution family.  A simulation using $N_{\rm gas} = 130536$
is added to study resolution effects.

In Fig.~\ref{jcoldstar}, we plot the integrated cooling and star
formation rates, the mass of the cold gas disc, and the specific
angular momentum of the cold gas disc.  The cooling rates are very
similar to those in the cooling only simulations, so star formation
does not greatly affect the cooling rate when a halo contains a
sufficient number of gas particles ($N_{\rm gas} \geq 5000$).  At the
end of the simulation there is a factor of two difference in the cold
gas masses as the resolution is varied. However, this is a small
fraction of the baryonic mass, with only 48, 94, 794, and 2192 cold
gas particles remaining in the $N_{\rm gas} = 5544, 15408, 44442$, and
130536 simulations, respectively.

As was seen in the cooling only simulations, the decline of the specific
angular momentum of the cold gas discs starts from $t \sim 3$ Gyr when the
rapid accretion of cold gas finishes. These decreases are
much stronger than those in the cooling only
simulations, because the cold gas discs now contain fewer
particles. Despite this, the two highest resolution simulations still
manage to produce reassuringly similar results.
Compared with the cooling only runs, the values of the specific
angular momenta of the cold gas discs are significantly higher. This is
because the low angular momentum cold gas is preferentially creamed off and
converted into stars.
 
\begin{figure}
\includegraphics[width=8cm]{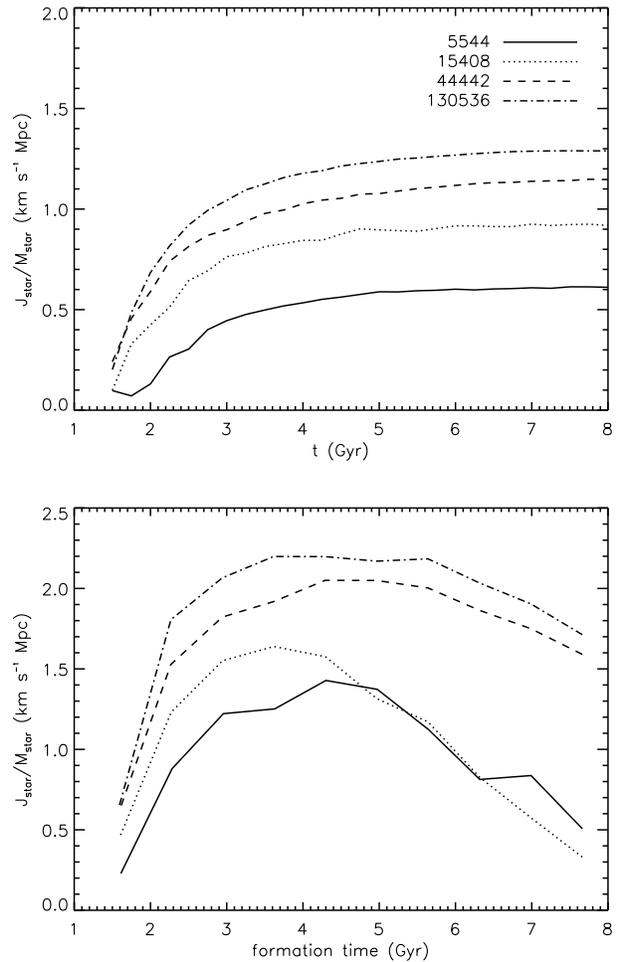}
\caption{
 Evolution of the stellar discs angular momentum (upper panel) and
 formation time-angular momentum distributions in the stellar discs at
 $t = 8$ Gyr (lower panel).  The solid, dotted, dashed, and dot-dashed
 lines indicate the simulations with $N_{\rm gas} = 5544, 15408,
 44442$, and 130536, respectively.  }
\label{jstar}
\end{figure}

The lower panel in Fig.~\ref{jstar} shows the final specific angular
momenta of the stars as a function of their formation time. The
overall shape of these curves reflects the angular momentum evolution
in the cold gas, albeit shifted downwards by $\sim 30$ per cent. The
declining stellar specific angular momentum at late times implies that
there would be outside-in disc formation, i.e. older stars have larger
angular momentum. The outside-in disc formation found by \citet{sgp02}
might be explained by this process.  It is crucial to obtain a
numerically robust estimate of the angular momentum evolution of the
cold gas before the stellar population distribution in the disc can be
reliably studied. 

The evolution of the total stellar specific angular momentum
is shown in the upper panel of Fig.~\ref{jstar}.
These results are dominated by the bulk of the stars
which form in the first few Gyr after cooling is switched
on. The small differences in the cold gas angular momenta for the two
highest resolution runs at these early times are sufficient to imprint
similar sized differences in the final stellar angular momenta.
It should be noted that the specific angular momenta of the stellar discs would
depend on the resolution even if the cold gas discs had exactly the
same specific angular momenta. 
The reason for this is that higher resolution enables higher density
regions to be resolved at large radii. In addition, a shorter gravitational
softening length gives higher gas density for a given surface gas density. 
Consequently, higher resolution allows higher angular momentum gas to
form stars. 
In order to achieve numerical convergence, we should include
self-regulated star formation tuned to give the surface density of the
star formation rate as a function of the surface gas density 
\citep[][see also Yepes et al. 1997; Hultman \& Pharasyn 1999]{gi97,
sh02b} as the observations suggest \citep{ken98}.
\citet{clc98} and \citet{buo00} have pointed out that inclusion of 
feedback processes, for example, Type Ia and II supernovae, stellar
winds, and ultraviolet radiation from massive stars, has a significant
impact on the evolution of model galaxies.  Including feedback
processes is, however, beyond the scope of this paper.

\begin{figure}
\includegraphics[width=8cm]{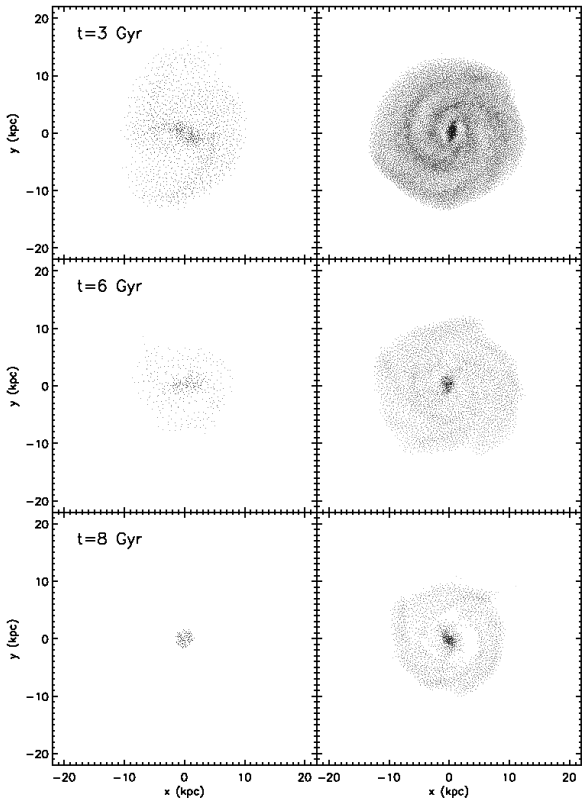}
\caption{Distribution of cold gas in the star formation runs,  
 $N_{\rm gas} = 15408$ (left)  and 130536 (right). 
 The grey scale is coded by the three dimensional gas density.  
}
\label{star2coldgas}
\end{figure}
Fig.~\ref{star2coldgas} shows the distribution of the cold gas
particles in the star formation runs with $N_{\rm gas} = 15408$ and
130536.  In the $N_{\rm gas} = 15408$ simulation, the cold gas disc
has some small holes at $t = 3$ Gyr. At $t = 6$ Gyr the holes have
become large and finally the cold gas disc shrinks to the centre at $t
= 8$ Gyr.  The evolution in the $N_{\rm gas} = 130536$ simulation
shows the same trend, but the disc is much smoother.  At $t = 3$ Gyr
the disc has beautiful spiral arms, and then some small holes appear
at $t = 6$ Gyr.  The final disc is much more extended than that in the
lower resolution simulation, but it has acquired unphysically large
holes.

\begin{table}
\caption{
 Hydrodynamic torques acting on the cold gas discs parallel to the
 angular momenta of cold gas discs at $t = 5.5$ Gyr.  The second,
 third, and fourth column show the total hydrodynamic torques, the
 hydrodynamic torques from the pressure gradient force, and the
 hydrodynamic torques caused by the artificial viscosity.  The torques
 are normalised by the angular momenta. The unit is Gyr$^{-1}$.  }
\label{torquenw93}
\begin{tabular}{@{}lccc}
\hline
$N_{\rm gas}$ & $\left(\frac{\tau_{\rm cold}}{J_{\rm cold}}\right)_{\rm hydro}$ & 
 $\left(\frac{\tau_{\rm cold}}{J_{\rm cold}}\right)_{\nabla P}$ & 
 $\left(\frac{\tau_{\rm cold}}{J_{\rm cold}}\right)_{\rm AV}$ \\ 
\hline
15408 (no SF) & -0.048 & -0.017 & -0.031 \\
28624 (no SF) & -0.035 & -0.013 & -0.021  \\
44442 (no SF) & -0.033 & -0.015 & -0.017 \\
\hline
15408 (SF) & -0.19 & -0.11 & -0.079 \\
44442 (SF) & -0.069 & -0.028 & -0.042 \\ 
130536 (SF) & -0.067 & -0.044 & -0.023 \\
\hline
\end{tabular}

\end{table}
Table.~\ref{torquenw93} displays the hydrodynamic torques acting on
the cold gas discs at $t = 5.5$ Gyr, both in the cooling-only and star
formation series.  We find that the cold gas discs in the simulations
with star formation received much stronger negative torques compared
with the cooling only runs with the same initial $N_{\rm gas}$.  Part
of the reason for this is that the torque from the artificial
viscosity has a strong dependence on the number of cold gas particles.
However, the contribution to the torque from the pressure gradients is
also significantly larger in the star formation runs. This is due to
ram pressure caused by the creation of holes once enough gas particles
have been converted to stars.  In the $N_{\rm gas} = 130536$
simulation, this torque becomes stronger than that caused by the
artificial viscosity.
Thus, even in this best-resolved simulation, the cold gas disc is
depleted to an extent where resolution-dependent holes are formed and
resolution-dependent fractions of the angular momentum are lost.

\section{DECOUPLING THE COLD PHASE FROM THE HOT PHASE}

In the previous sections we have seen that the angular momentum
transfer from the cold gas to the hot halo gas is mainly caused by
numerical problems which are intrinsic to the SPH technique.  The
collapsing rotating sphere test in the previous section reveals that
the problem becomes worse when star formation is included.  The reason
for this is as follows.
Star formation decreases the number of particles in the cold gas disc
causing the disc to becomes thinner.  We had found in the preceding section
that a disc is prone to develop holes when the number of particles in
it is not sufficient.
Once these holes appear, ram pressure between the hot and cold gas
phases becomes important and a rapid loss of angular momentum from the
cold gas ensues.

It is interesting therefore to see what happens when we inhibit the
angular momentum transfer from cold gas to hot gas in simulations of
galaxy formation.  This can be achieved in a drastic way by decoupling
the cold and hot phases.  \citet{pea99} proposed a decoupling
technique whereby the contribution to the hot gas particles' density
from the cold gas particles is explicitly ignored. This has the effect
of suppressing the overcooling of the hot gas that otherwise
results. In this scheme hot and cold particles nonetheless interact
through mutual hydrodynamical forces.  Our approach is more radical:
hot and cold particles do not interact at all except through gravity.
The cold and hot phases are treated as independent fluids.  Our
approach is quite similar to the multi-phase model proposed by
\citet{sc02} in which they represents a warm phase by SPH particles
and the ISM by sticky particles and do not allow the ISM to interact
with ambient gas hydrodynamically.  While clearly we miss some real
physics as a result of such decoupling, the physics that we miss
cannot be modelled properly using SPH anyway as we have shown.  In the
following subsections we show simulations of disc formation adopting
this decoupling technique.

\subsection{Disc formation in a rotating sphere}

\begin{figure*}
\includegraphics[width=16cm]{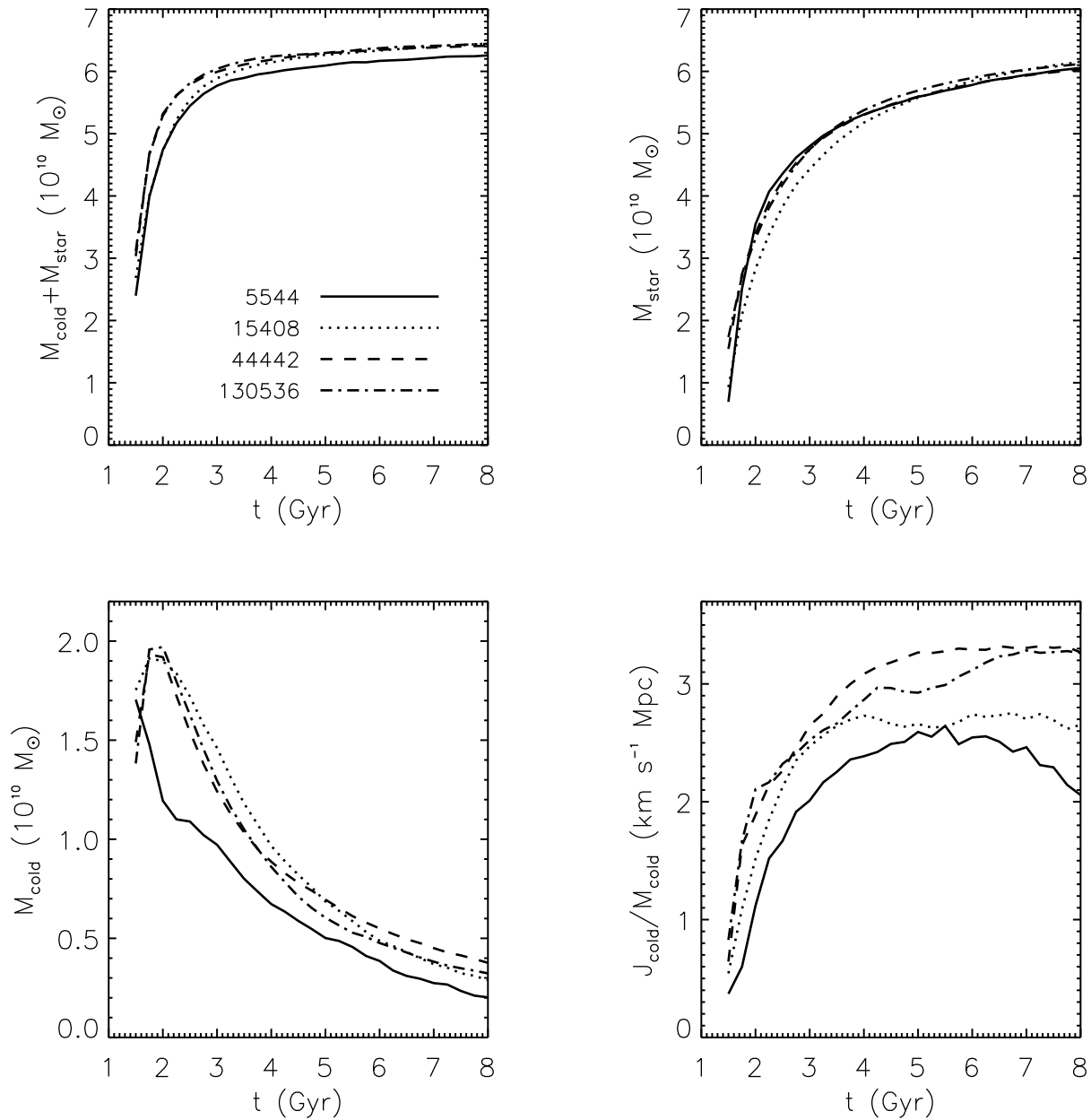}
\caption{
	The same as Fig.~\ref{jcoldstar} but with the decoupling. 
}
\label{jcolddcp}
\end{figure*}

The technique of decoupling is appropriate in situations in which the
pressure from the hot gas is unimportant in confining the cold gas
component and when the pressure at the midplane of the cold gas disc
is sufficiently greater than the external pressure on the contact
surfaces. In such a situation, the self-gravitating gas disc can be a
good approximation.  We now present simulations including star
formation, which are similar to those discussed in Section~4.2.2 and
illustrated in Figs~14-16.

Fig.~\ref{jcolddcp} shows the evolution of the cold gas disc.  The
integrated cooling rates are almost the same as in
Fig.~\ref{jcoldstar}.  The stellar mass is also similar to that in the
normal simulation, but more mass remains as cold gas (about a factor
of 2).  The evolution of the specific angular momenta of the cold gas
discs are quite different from those in Fig.~\ref{jcoldstar}. Apart
from the lowest resolution simulation, the simulations no longer show
a decrease in the specific angular momentum as a function of
time. Even the lowest resolution simulation shows a decline which
starts much later and the disc ends up with much larger specific
angular momentum than in the normal SPH simulation.  This proves that
most of the angular momentum loss from the cold gas disc is due to the
hydrodynamic interaction with the ambient hot gas, which cannot be
dealt with adequately by SPH.  A resolution dependence is still
present, because the accreting gas can have large velocity shears
before it reaches the temperature below which it becomes decoupled
from the hot gas.  We have also investigated the torque that causes
the decline of the angular momentum of the cold gas disc in the lowest
resolution simulation.  We find that most of the negative torque is
coming from the gravitational interaction with stellar and dark matter
particles.  We conjecture that this is because the density
distributions of the cold gas disc and the stellar disc are not smooth
enough to prevent angular momentum transfer due to tidal torques.

\begin{figure}
\includegraphics[width=8cm]{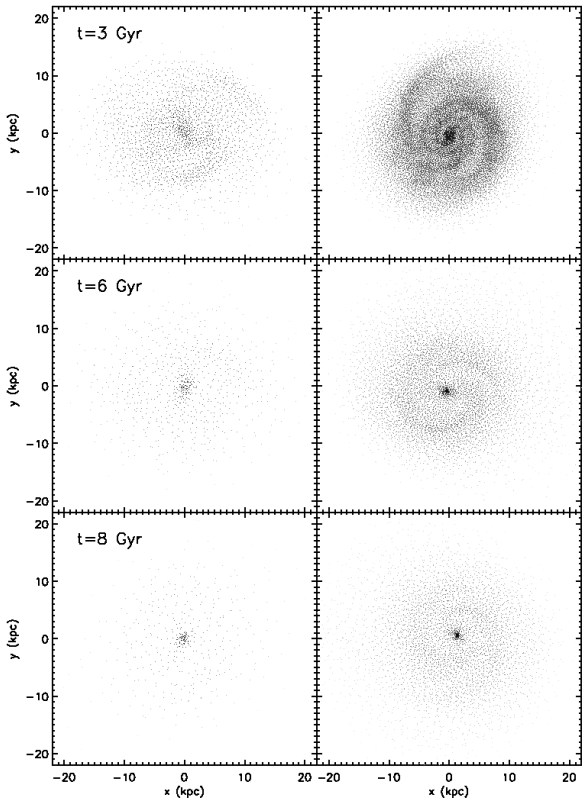}
\caption{
The same as Fig.~\ref{star2coldgas} but with the decoupling.  }
\label{stardcpcoldgas}
\end{figure}

In Fig.~\ref{stardcpcoldgas}, we show the face-on views of the cold
gas discs in the simulations using $N_{\rm gas} = 15408$ and 130536.
Both produce smooth extended cold gas discs without any holes. The
main difference due to resolution is in the three-dimensional density
of cold gas particles and in the ability to resolve spiral arms. These
are direct consequences of the differences in the spatial resolution
of gravity and mass resolution in the SPH.

\begin{figure}
\includegraphics[width=8cm]{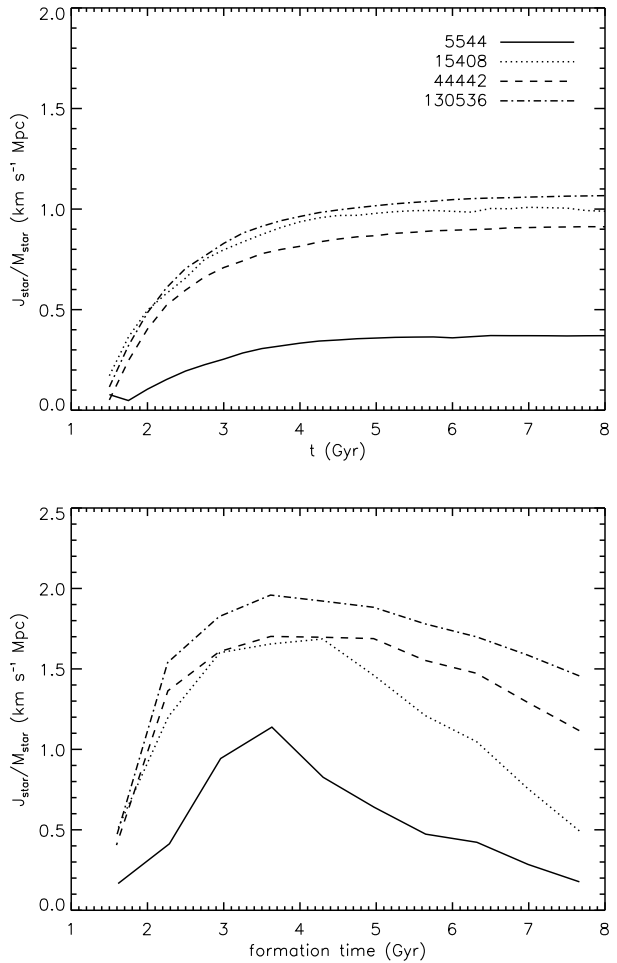}
\caption{
	The same as Fig.~\ref{jstar} but with decoupling.  }
\label{jstardcp}
\end{figure}

The evolution of the stellar discs and their age-angular momentum
distribution at $t = 8$ Gyr are presented in Fig.~\ref{jstardcp}.
Except for the lowest resolution simulation, the angular momentum
evolution shows better convergence than in the normal SPH
simulations. The reason why the stellar disc in the $N_{\rm gas} =
15408$ simulation has larger specific angular momentum than that in
the $N_{\rm gas} = 44442$ simulation is as follows.  Although the cold
phase is decoupled from the hot phase, angular momentum transfer to
the hot gas during accretion before the gas temperature has reached
the decoupling threshold ($1.3 \times 10^5$~K; see Section~4.2.1) is
still allowed.  Consequently, the lower resolution simulation produces
a lower angular momentum cold gas disc.  Since the local star
formation rate is a function of density and we impose a threshold
density for star formation, the low angular momentum of the cold gas
disc does not always result in a lower angular momentum for the
stellar disc.  Higher angular momentum gas particles are often not
dense enough to be eligible for star formation.  Thus, the angular
momentum transfer that brings such gas particles to the inner disc
where the gas particles can form stars sometimes produces a higher
angular momentum stellar disc.  The increasing resolution is also
likely to increase the angular momentum of the stellar disc, because
higher mass resolution allows fragmentation to be resolved in less 
dense environments. As a result, the high angular momentum
gas particles that cannot form stars in a low resolution simulation
are allowed to form stars if they reside in the dense structures like
spiral arms that are resolved with higher resolution (see
Fig.~\ref{stardcpcoldgas}).

The above complex picture also provides a reasonable explanation for
the fact that, at the highest resolution, the stellar disc in the
normal SPH simulation (Fig~15) has a slightly higher angular momentum
than in the simulation with decoupling (Fig~19), despite the fact that
the angular momentum of the cold gas disc in the normal SPH simulation
is lower than in the simulation with decoupling.  The angular momentum
transfer from the cold gas makes more cold gas particles eligible to
form stars and the holes in the cold gas disc enhance the density of
the cold gas disc at large radii.  Of course, the pressure from the
hot gas also enhances the density and thus the star formation rate all
over the disc, but, as we have seen, this process cannot be modelled
appropriately by SPH.

The age-angular momentum distribution exhibits the outside-in disc
formation feature again, although this is weaker than in the standard
SPH simulation.  This is not surprising because the star forming
region shrinks when the surface density of the cold gas disc is
reduced by star formation. However, as we mentioned above, higher
resolution allows stars to form in the outskirts of the disc even when
the surface gas density becomes quite low. Hence the outside-in
feature becomes weaker with increasing resolution. Even though the
resolution limits for the $N_{\rm gas} = 44442$ and 130536 simulations
(Eq. \ref{jeans2}) are far below the threshold density for star
formation ($\rho_{\rm th} = 10^{-25}$ g cm$^{-3}$), the age-angular
momentum distributions do not converge.  Self-regulated star formation
may remove this resolution dependence as we discussed previously.  We
will test this in future work.

\subsection{Cosmological simulations}

Finally, we present cosmological simulations adopting the decoupling
of the cold and hot phases. In this case, we allow warm gas ($3
\times 10^4 < T < 5 \times 10^5$ K) to interact with both the cold and
hot phases so that the simulations do not fail when there are too few
cold gas particles to perform the SPH calculation.  All conditions are
the same as the simulations presented in Section
\ref{cos1} except for the decoupling between the hot and cold
phases.

\begin{figure}
\includegraphics[width=8cm]{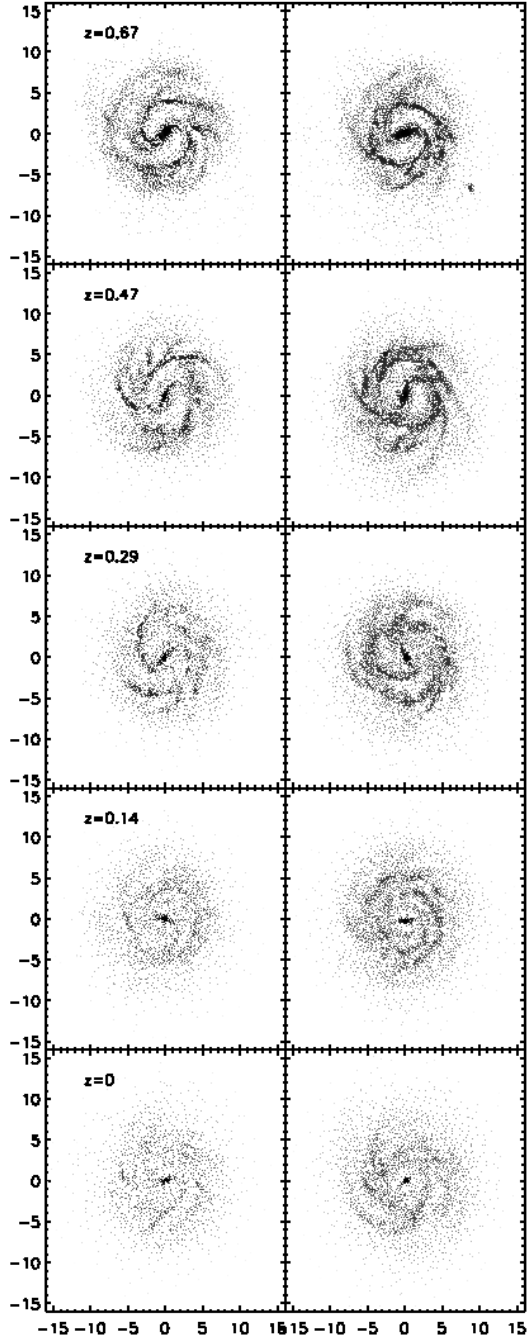}
\caption{Face-on views of the distribution of cold gas. 
 the left panels show the galaxy in the simulation with decoupling and
 the ``conversion'' star formation prescription and the right panels
 for the simulation with decoupling and the ``spawning'' star
 formation prescription.  The length is in units of $h^{-1}$ kpc. }
\label{redshift2}
\end{figure}
In Fig.~\ref{redshift2}, we show the redshift evolution of the cold
gas discs in the decoupling simulations. As in Section \ref{cos1}, we
adopt two star formation schemes, ``conversion'' and ``spawning.''
The distributions of the cold gas particles are quite different from
those in the standard SPH simulations (see Fig.~\ref{z0} and
Fig.~\ref{redshift}).  In the standard SPH simulations, most of the
cold gas particles are in the filaments and the remaining regions are
almost empty. Now the galaxies have smooth, extended cold gas
discs. These gas discs have spiral arms instead of filaments.  Since
the cold gas particles have a temperature $\sim 10^4$~K, that is one
order of magnitude smaller than the lower limit in the idealised
simulations of Section~4, the cold gas discs are much thiner.  This
makes the problems encountered in the shear tests more serious.
Consequently, the cold gas disc will have an unphysical morphology
because of numerical effects {\it unless} we decouple the cold gas
from the hot halo gas.  Note that the smooth density distribution of
the cold gas disc significantly decreases angular momentum transfer
due to tidal torques as well.

\begin{figure}
\includegraphics[width=7.5cm]{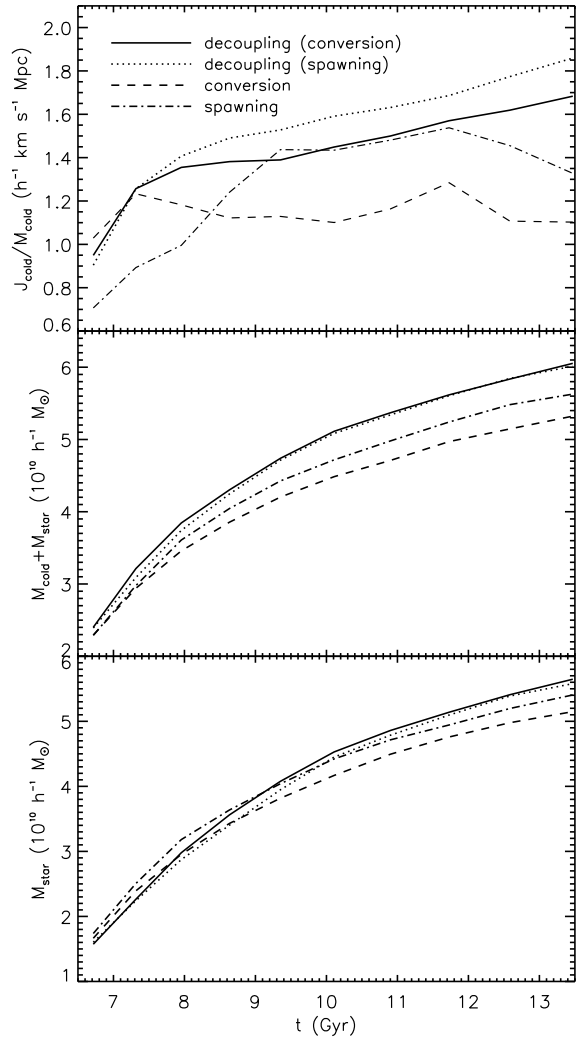}
\caption{
Evolution of the galaxies. The top, middle, and bottom panels present
the specific angular momenta of the cold gas discs, the integrated
cooling rates, and the integrated star formation rates, respectively.
The solid and dotted lines indicate the decoupling simulations with
the conversion and spawning star formation recipes, respectively.  The
results from the standard SPH simulations are also shown by the dashed
(conversion) and dot-dashed (spawning) lines for reference.  }
\label{jgasdcp}
\end{figure}

Fig.~\ref{jgasdcp} shows the specific angular momenta of the cold gas
discs, the integrated cooling rates, and the integrated star formation
rates in the galaxies as functions of time.  The results from the
standard SPH simulations (Fig.~3) are also plotted for reference.
Now, with decoupling, the specific angular momenta of the cold gas
discs become monotonically increasing functions of time.  There is a
small difference between the specific angular momenta of the cold gas
discs in the two decoupling simulations.  This might be because the
cold gas disc in the spawning case is less affected by the angular
momentum transfer between the cold gas and the warm gas, although we
cannot find any significant difference in the hydrodynamic and tidal
torques in these two simulations.  To decide whether the high specific
angular momentum of the cold gas disc in the spawning run is the
result of the larger number of cold gas particles or a side effect of
the multi-mass SPH imposed by this star formation prescription, we
would have to perform another conversion simulation with a larger
number of the SPH particles leaving all other conditions unchanged.
In this paper we do not perform this test because it would require a
much larger number of particles in order to satisfy the Jeans
condition (Eq. \ref{jeans2}) beyond the threshold density for star
formation. 

The cooling rates and the star formation rates are almost the same
between the two decoupling simulations, and they are higher than in
the standard SPH simulations. This proves that the numerical angular
momentum feedback puffs up the hot halo gas and the cooling rate is
reduced in the standard SPH simulations.  Because this effect was not
observed in the idealised simulations, we conclude that as the
temperature of the cold gas disc becomes lower, the numerical angular
momentum transfer becomes more problematic.

\begin{figure}
\includegraphics[width=7.5cm]{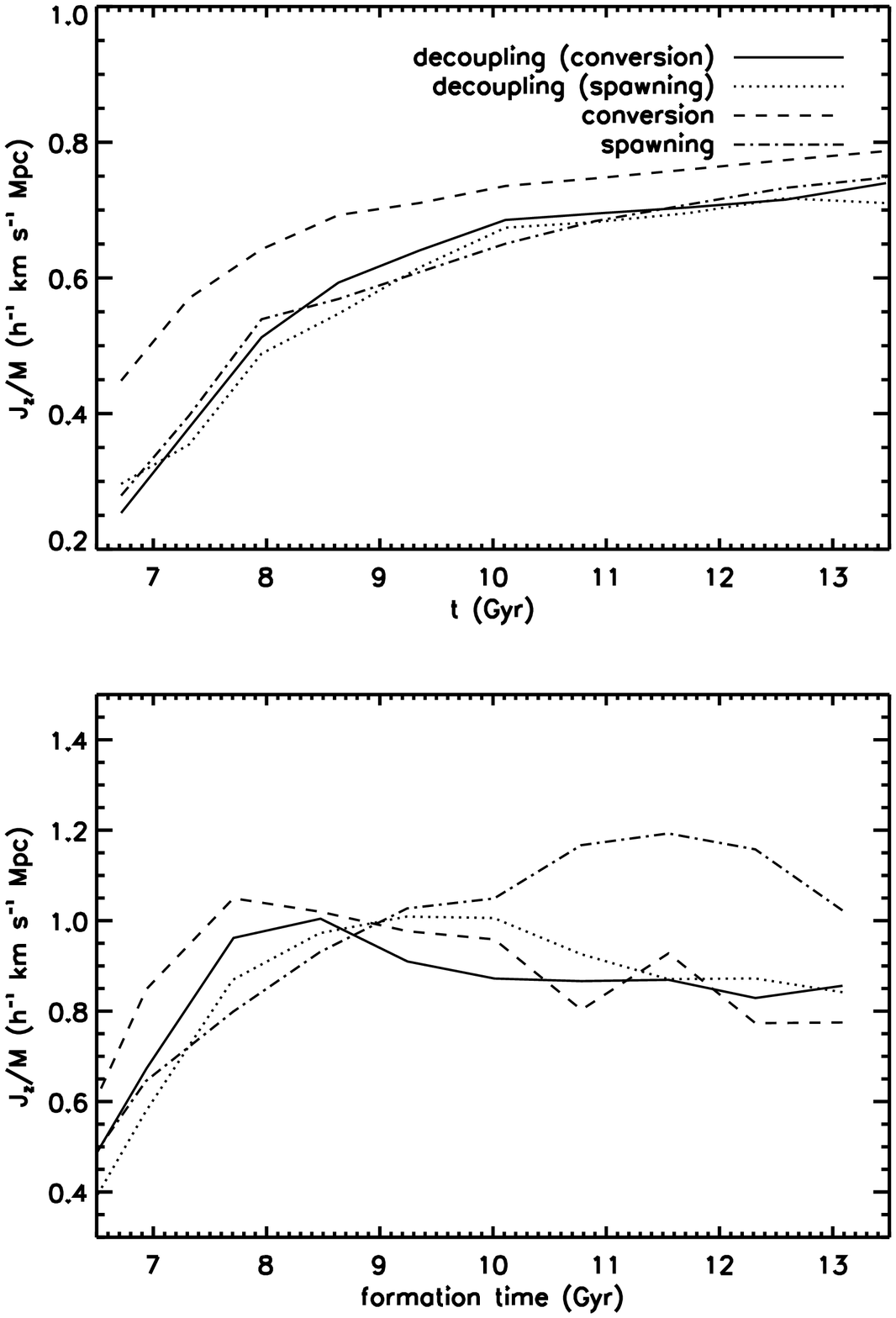}
\caption{
 Evolution of the specific angular momenta of stellar discs as a
 function of time and the age-angular momentum distributions of stars
 at $z = 0$.  The solid, dotted, dashed, and dot-dashed lines indicate
 the decoupling simulations with conversion and spawning star
 formation and the standard SPH simulations with conversion and
 spawning star formation, respectively.  }
\label{jstarcos}
\end{figure}
To find out how the angular momentum transfer affects the stellar
disc, we plot the evolution of the specific angular momenta of the
stellar discs and the age-angular momentum distributions at $z = 0$ in
the decoupling and standard SPH simulations in Fig.~\ref{jstarcos}.
Interestingly, except for the standard conversion simulation, all the
other simulations show similar evolution of the specific angular
momentum of the stellar discs.  On the other hand, except for the
standard spawning simulation, all the simulations produce similar
results for the age-angular momentum distributions.  These may be just
a coincidence.  Since we impose a relatively high density threshold as
a star formation criterion, only cold gas particles that have small
angular momenta can form stars and thus the threshold density
determines the angular momenta of the stellar discs.  It makes the
specific angular momentum evolution of the stellar discs almost
independent of those of the cold gas discs.

In the standard SPH simulation with the conversion star formation
prescription, the formation of holes in the cold gas disc results in
the cold gas being swept out to form a dense ring at large radius at
high redshifts (see the top left panel of Fig.~\ref{redshift}).  Gas
particles having large angular momenta can form high angular momentum
stars in this ring.  On the other hand, in the standard spawning
simulation, the gas disc has a large ring at low redshift (see the
bottom right panel of Fig.~\ref{redshift}).  From this ring, the high
angular momentum population stars of age $\sim 2$ Gyr are born.  Since
these holes are numerical artifacts as we showed in the shear tests,
the large angular momentum of the stellar disc and the high angular
momentum population in the standard SPH simulations are numerical
artifacts as well.  If the gas disc is puffed up by some feedback
processes and if we choose a lower threshold density for star
formation, the simulations with and without decoupling would produce
different results as in the case of the idealised simulations with the
temperature floor at $T = 10^5$ K.  The idealised simulations and the
cosmological simulations presented in this paper suggest that the
angular momentum transfer from the cold gas disc sometimes results in
too large and sometimes too small angular momentum of the stellar disc
depending on the resolution and the modelling of the ISM and star
formation.  It thus must depend on the modelling of feedback as well.

\section{SUMMARY AND DISCUSSION}

We have found in a cosmological SPH simulation that a disc of cold gas
breaks up into filaments and there is a significant transfer of
angular momentum from the cold gas disc to the ambient hot halo gas.
The dominant contribution to the hydrodynamical torque between the
cold and hot phases comes from the pressure gradient forces and not
from the artificial viscosity. The hot gas is puffed up by this
angular momentum `feedback' and this, in turn, can affect the cooling
rate.

By using simple shear tests, we find that SPH cannot correctly solve
problems where there are strong shear flows.  When a dense cold gas
sheet is moving in ambient diffuse hot gas, the gas sheet receives
strong negative acceleration due to pressure gradients and artificial
viscosity. By varying the number of neighbouring particles involved in
the SPH smoothing, we find that the deceleration due to pressure
gradients is related to the noisiness in SPH variables.  Moreover, if
the cold gas sheet does not contain a sufficient number of particles,
it undergoes hole formation.  Ram pressure from the hot
gas in these holes leads to further momentum transfer.

The comparison of simulations using SPH and a finite-difference
Eulerian code clearly exhibits a shortcoming of the SPH method: SPH
does not generate the turbulent mixing of the fluid components in the
shear flow tests that is present with the Eulerian code. Instead, with
SPH the fluid velocities change to damp out the shear flow but without
any mixing of fluids.  This feature of SPH only becomes problematic
when there are large velocity gradients.  Since previous test
simulations did not consider such a large shear, SPH has produced
promising results.  However, as we have shown in this paper, such a
large shear can exist in galaxy formation problems where a
rotationally supported cold gas disc forms within hot halo gas that is
mainly supported by thermal pressure.

The idealised simulations of disc formation in rotating hot gas reveal
that this problem has a strong dependence on the resolution.  Having
star formation in a simulation makes the situation worse by decreasing
the number of cold gas particles. If not enough particles are used,
the angular momentum of the cold gas disc steeply declines with time.
Consequently, the stellar disc forms from the outside to the inside.
This process may explain the outside-in formation of disc galaxies
found in the simulations by \citet{sgp02}.

One way to avoid these problems, is to decouple the cold and hot gas
phases. In the simulations with decoupling, the specific angular
momenta of the cold gas discs become monotonically increasing
functions of time as expected on theoretical grounds. The main
difference from the standard SPH simulations is seen in the morphology
of cold gas discs.  The simulations with decoupling produce smooth
extended cold gas discs that do not have any holes.  These simulations
also reveal that the numerical angular momentum transfer sometimes
increases the specific angular momenta of stellar discs and sometimes
decreases them, while it always decreases the specific angular momenta
of cold gas discs.  This strange feature is caused by the density
dependence of star formation. Hence, the way in which this problem
affects stellar discs is highly dependent on resolution and on the
modelling of subgrid physics in the interstellar medium.

We do not believe that the angular momentum transfer that we have
identified here is the main explanation for the angular momentum
problem in galaxy formation simulations, because the difference in the
specific angular momenta of stellar discs between simulations with and
without decoupling is not that large, although it seems to depend
strongly on the modelling of subgrid physics such as star formation.
At higher resolution, the numerical breaking up of cold gas discs is
likely to increase the specific angular momentum of the stellar discs
by enhancing the cold gas density and hence star formation rate in the
outskirts of the discs.  Anyway, as long as the cold gas disc suffers
spurious angular momentum transfer and has a strange morphology, the
properties of the resulting stellar disc are quite unreliable.

The angular momentum transfer rakes up all the cold gas into the star
forming region and spurious fragmentation of the disc induces quite
effective star formation in these dense filaments.  Consequently, the
problem can be much more serious when one investigates the details of
simulated galaxies like the cold gas distribution, the hot gas
distribution, the distribution of stellar populations, and all
observables related to them.  The problem may also affect simulations
of elliptical galaxies.  After the galaxy has exhausted most of its
cold gas in, for example, a starburst, newly accreting cold gas is
quickly supplied to the centre regardless of its angular momentum
(newly accreting cold gas implies the existence of halo gas).  The
decoupling of the cold and hot gas phases that we have introduced
offers the opportunity to investigate the detailed structure of
galaxies by avoiding this spurious angular momentum transfer.

However, it seems clear that complex physical processes taking place
in the interstellar medium and in the hot halo gas must play a key
role in galaxy formation.  A code that can solve problems involving
large shear motions, together with the ability to treat a wide dynamic
range, is required to study these processes in detail.  AMR is an
obvious candidate, although it has not yet been widely used in this
subject, and thus still needs substantial testing.  On the other hand,
substantial refinements of SPH have been introduced recently which
could prove useful in this context, for example
\citet[][redefined particle velocity by locally averaged velocity]{mon89}, 
\citet[][tensorial smoothing kernels]{owe98}, 
\citet[][smoothed pressure SPH]{rt01}, 
\citet[][SPH with Riemann solver]{inu02}, 
\citet[][consistent particle velocity with fluid velocity]{ii02}, 
\citet[][adaptive mass resolution]{kw02}, and 
\citet[][conservation of both entropy and energy]{sh02}. 
We encourage colleagues who have developed and implemented these
refinements to perform the shear tests presented in this paper.  The
implementation by \citet{ii02} is particularly interesting since it
substantially suppresses spurious density errors in SPH calculations
of shear flows. 

The phase decoupling technique that we have introduced may be regarded
as a crude way of modelling a multi-phase fluid in cosmological
simulations and seems to produce much better results than standard
SPH. This decoupling seems a reasonable approximation when the disk is
self-gravitating and consists of cold gas and stars. As the
multi-phase structure of the interstellar medium begins to be resolved
in a simulation, this approximation must break down since the external
pressure from halo gas plays a role in confining the hot phase of the
interstellar medium. Modelling these processes remains a challenge
which may hold the key for realistic simulations of the formation of
galactic discs. 

\section*{Acknowledgments}

We are grateful to Volker Springel for kindly providing the improved
version of GADGET.  We also thank to Simon White, Lars Hernquist,
Peter Thomas, Julio Navarro, Rob Thacker and Richard Bower for their
highly useful comments on this work.  We acknowledge the financial
support from UK PPARC.  VRE acknowledges a Royal Society University
Research Fellowship.  VQ is a Ramon y Cajal Fellow from the Spanish
Ministry of Science and Technology and has partial financial support
form grant AYA2000-2045.



\bsp

\label{lastpage}


\begin{thebibliography}{99}
\bibitem[\protect\citeauthoryear{Abel, Bryan \& Norman}{2000}]{abn00}
Abel T., Bryan G. L., Norman M. L., 2000, ApJ, 540, 39
\bibitem[\protect\citeauthoryear{Bate \& Burkert}{1997}]{bb97} 
	Bate M. R., Burkert A., 1997, MNRAS, 288, 1060
\bibitem[\protect\citeauthoryear{Balsara}{1995}]{bal95} 
	Balsara D. W., 1995, J. Comp. Phys., 121, 357
\bibitem[\protect\citeauthoryear{Buonomo et al.}{2000}]{buo00} 
        Buonomo F., Carraro G., Chiosi C., Lia C., 2000, MNRAS, 312,
        371
\bibitem[\protect\citeauthoryear{Carraro, Lia \& Chiosi}{1998}]{clc98} 
         Carraro G., Lia C., Chiosi C., 1998, MNRAS, 297, 1021
\bibitem[\protect\citeauthoryear{Cole et al.}{2000}]{col00} 
	Cole S., Lacey C. G., Baugh C. M., Frenk C. S., 2000, MNRAS,
	319, 168
\bibitem[\protect\citeauthoryear{Dalcanton, Spergel \& Summers}{1997}]{dss97} 
	Dalcanton J. J., Spergel D. N., Summers F. J., 1997, ApJ, 482
\bibitem[\protect\citeauthoryear{Davis et al.}{1985}]{defw85} Davis,
M., Efstathiou, G., Frenk, C. S., White, S. D. M., 1985, ApJ, 292, 371
659
\bibitem[\protect\citeauthoryear{Eke, Efstathou \& Wright}{2000}]{eew00} 
	Eke V. R., Efstathiou G., Wright L., 2000, MNRAS, 315, L18
\bibitem[\protect\citeauthoryear{Evrard}{1988}]{evr88} 
	Evrard A. E., 1988, MNRAS, 235, 911
\bibitem[\protect\citeauthoryear{Evrard, Summers \&
         Davis}{1994}]{esd94} Evrard, A. E., Summers, F. J., Davis, M.
         1994, ApJ, 422, 11
\bibitem[\protect\citeauthoryear{Fall \& Efstathiou}{1980}]{fe80} 
	Fall S. M., Efstathiou G., 1980, MNRAS, 193, 189
\bibitem[\protect\citeauthoryear{Frenk et al} {1996}]{fews96} Frenk,
        C. S., Evrard, A. E., White, S. D. M., Summers, F. J., 1996,
        ApJ, 472, 460
\bibitem[\protect\citeauthoryear{Gerritsen \& Icke}{1997}]{gi97} 
	Gerritsen J. P. E., Icke V., 1997, A\&A 325, 972
\bibitem[\protect\citeauthoryear{Governato et al.}{2002}]{gov02} 
	Governato F., Mayer L., Wadsley J. P., Gardner J. P., Willman
	B., Hayashi E., Quinn T., Stadel J., Lake G., 2002, preprint
	(astro-ph/0207044)
\bibitem[\protect\citeauthoryear{Haardt \& Madau}{1996}]{hm96} 
	Haardt F., Madau P., 1996, ApJ, 461, 20
\bibitem[\protect\citeauthoryear{Haltman \& K\"{a}llander}{1997}]{hk97} 
	Haltman J., K\"{a}llander D., 1997, A\&A, 324, 534
\bibitem[\protect\citeauthoryear{Hernquist \& Katz}{1989}]{hk89}
	Hernquist L, Katz N., 1989, ApJ, 70, 419
\bibitem[\protect\citeauthoryear{Hultman \& Pharasyn}{1999}]{hp99}
	Hultman J., Pharasyn A., 1999, A\&A, 347, 769
\bibitem[\protect\citeauthoryear{Imaeda \& Inutsuka}{2002}]{ii02} 
	Imaeda Y., Inutsuka S., 2002, ApJ, 565, 501
\bibitem[\protect\citeauthoryear{Inutsuka}{2002}]{inu02} 
	Inutsuka S., 2002, J. Comp. Phys., 179, 238
\bibitem[\protect\citeauthoryear{Katz}{1992}]{kat92} 
	Katz N., 1992, ApJ, 391, 502
\bibitem[\protect\citeauthoryear{Katz, Hernquist \& Weinberg}{1992}]{khw92} 
	Katz N., Hernquist L., Weinberg D. H., 1992, ApJ, 399, L109
\bibitem[\protect\citeauthoryear{Katz, Weinberg \& Hernquist}{1996}]{kwh96} 
	Katz N., Weinberg D. H., Hernquist L., 1996, ApJS, 105, 19
\bibitem[\protect\citeauthoryear{Kauffmann, White \& Guiderdoni}{1993}]{kau93}
	Kauffmann G., White S. D. M., Guiderdoni B., 1993, MNRAS, 264,
	201
\bibitem[\protect\citeauthoryear{Kitsionas \& Whitworth}{2002}]{kw02} 
	Kitsionas S., Whitworth A. P., 2002, MNRAS, 330, 129
\bibitem[\protect\citeauthoryear{Kennicutt}{1998}]{ken98} 
	Kennicutt R. C., 1998, ApJ, 498, 541
\bibitem[\protect\citeauthoryear{Landau \& Lifshitz}{1987}]{ll87} 
	Landau L. D., Lifshitz E. M., 1987, Course in Theoretical
	Physics: Fluid Mechanics, Institute of Physical Problems, USSR
	Academy of Sciences, Moscow
\bibitem[\protect\citeauthoryear{Lombardi et al.}{1999}]{lom99} 
	Lombardi J. C., Sills A., Rasio F. A., Shapiro S. L., 1999,
	J. Comp. Phys., 152, 687
\bibitem[\protect\citeauthoryear{Mo, Mao \& White}{1998}]{mmw98} 
	Mo H. J., Mao S., White S. D. M., 1998, MNRAS, 295, 319
\bibitem[\protect\citeauthoryear{Monaghan}{1985}]{mon85} 
	Monaghan J. J., 1985, J. Comp. Phys. Rep, 3, 71
\bibitem[\protect\citeauthoryear{Monaghan}{1989}]{mon89} 
	Monaghan J. J., 1989, J. Comp. Phys., 82, 1
	Monaghan J. J., 1997, J. Comp. Phys., 136, 298
\bibitem[\protect\citeauthoryear{Monaghan \& Gingold}{1983}]{mg83} 
	Monaghan J. J., Gingold R. A., 1983, J. Comp. Phys., 52, 374
\bibitem[\protect\citeauthoryear{Nagashima et al.}{2001}]{nag01} 
	Nagashima M., Totani T., Gouda N., Yoshii Y., 2001, ApJ, 557,
	505
\bibitem[\protect\citeauthoryear{Navarro \& White}{1993}]{nw93} 
	Navarro J. F., White S. D. M., 1993, MNRAS, 319, 619
\bibitem[\protect\citeauthoryear{Navarro \& White}{1994}]{nw94} 
	Navarro J. F., White S. D. M., 1994, MNRAS, 267, 401
\bibitem[\protect\citeauthoryear{Navarro, Frenk \& White}{1995}]{nfw95} 
	Navarro J. F., Frenk C. S., White S. D. M., 1995, MNRAS, 275,
	56
\bibitem[\protect\citeauthoryear{Nelson \& Papaloizou}{1994}]{np94} 
	Nelson R. P., Papaloizou, J. C. B., 1994, MNRAS, 270, 1
\bibitem[\protect\citeauthoryear{Netterfield et al.}{2002}]{net02} 
	Netterfield et al., 2002, ApJ, 571, 604
\bibitem[\protect\citeauthoryear{Okamoto \& Nagashima}{2003}]{on03} 
	Okamoto T., Nagamashima M., 2003, ApJ, 587, 500
\bibitem[\protect\citeauthoryear{Owen et al.}{1998}]{owe98} 
	Owen J. M., Vllumsen J. V., Shapiro P. R., Martel H., 1998,
	ApJS, 115, 155
\bibitem[\protect\citeauthoryear{Pagels \& Primack}{1982}]{pp82} 
	Pagels H., Primack J. R., 1982, Phys. Rev. Lett., 48, 223
\bibitem[\protect\citeauthoryear{Pearce et al.}{1999}]{pea99} 
	Pearce F. R. et al., 1999, (The Virgo Consortium), ApJ, 521,
	L99
\bibitem[\protect\citeauthoryear{Quilis, Ib\'{a}\~{n}ez \&
S\'{a}ez}{1996}]{qis96} Quilis V., Ib\'{a}\~{n}ez J. M., S\'{a}ez D.,
1996, ApJ, 469, 11
\bibitem[\protect\citeauthoryear{Quilis, Moore \& Bower}{2000}]{qmb00} 
	Quilis V., Moore B., Bower R. G., 2000, Sci, 288, 1617
\bibitem[\protect\citeauthoryear{Quilis, Bower \& Balogh}{2001}]{qbb01} 
	Quilis V., Bower R. G., Balogh M. L., 2001, MNRAS, 328, 1091
\bibitem[\protect\citeauthoryear{Rasio \& Shapiro}{1991}]{rs91} 
	Rasio F. A., Shapiro S. L., 1991, ApJ, 377, 559
\bibitem[\protect\citeauthoryear{Ritchie \& Thomas}{2001}]{rt01} 
	Ritchie B. W., Thomas P. A., 2001, MNRAS, 323, 743
\bibitem[\protect\citeauthoryear{Sch\"{u}ssler \& Schmitt}{1981}]{ss81} 
	Sch\"{u}ssler M., Schmitt D., 1981, A\&A, 97, 373
\bibitem[\protect\citeauthoryear{Semelin \& Combes}{2002}]{sc02} 
	Semelin B., Combes F., 2002, A\&A, 388, 829
\bibitem[\protect\citeauthoryear{Somerville \& Primack}{1999}]{sp99} 
	Sommerville R. S., Primack J. R., 1999, MNRAS, 310, 1087
\bibitem[\protect\citeauthoryear{Sommer-Larsen \& Dolgov}{2001}]{sd01} 
	Sommer-Larsen J., Dolgov A., 2001, ApJ, 551, 608
\bibitem[\protect\citeauthoryear{Sommer-Larsen, G\"{o}tz \&
Portinari}{2002}]{sgp02} Sommer-Larsen J., G\"{o}tz M., Portinari L.,
2002, preprint (astro-ph/0204366)
\bibitem[\protect\citeauthoryear{Springel}{2000}]{spr00} 
	Springel V., 2000, MNRAS, 307, 162
\bibitem[\protect\citeauthoryear{Springel \& Hernquist}{2002}]{sh02} 
	Springel V., Hernquist L., 2002, MNRAS, 333, 649
\bibitem[\protect\citeauthoryear{Springel \& Hernquist}{2003}]{sh02b} 
	Springel V., Hernquist L., 2003, MNRAS, 339, 289
\bibitem[\protect\citeauthoryear{Springel, Yoshida \& White}{2001}]{syw01} 
	Springel V., Yoshida N., White S. D. M., 2001, New Astronomy,
	6, 79
\bibitem[\protect\citeauthoryear{Steinmetz}{1996}]{ste96} 
	Steinmetz M., 1996, MNRAS, 278, 1005
\bibitem[\protect\citeauthoryear{Steinmetz \& M\"{u}ller}{1993}]{sm93} 
	Steinmetz M., M\"{u}ller E., 1993, A\&A, 268, 391
\bibitem[\protect\citeauthoryear{Steinmetz \& Navarro}{1999}]{sn99} 
	Steinmetz M., Navarro J. F., 1999, ApJ, 513, 555
\bibitem[\protect\citeauthoryear{Steinmetz \& Navarro}{2002}]{sn02} 
	Steinmetz M., Navarro J. F., 2002, New Ast., 7, 155
\bibitem[\protect\citeauthoryear{Sutherland \& Dopita}{1993}]{sd93} 
	Sutherland R. S., Dopita M. A., 1993, ApJS, 88, 253
\bibitem[\protect\citeauthoryear{Thacker \& Couchman}{2000}]{tc00} 
	Thacker R. J., Couchman H. M. P., 2000, ApJ, 545, 728
\bibitem[\protect\citeauthoryear{Thacker \& Couchman}{2001}]{tc01} 
	Thacker R. J., Couchman H. M. P., 2001, ApJ, 555, L17
\bibitem[\protect\citeauthoryear{Thacker et al.}{2000}]{tha00} 
	Thacker R. J., Tittley E. R., Pearce F. R., Couchman H. M. P.,
	Thomas P. A., 2000, MNRAS, 319, 619
\bibitem[\protect\citeauthoryear{Theuns et al.}{1998}]{the98} 
	Theuns T., Leonard A., Efstathiou G., Pearce F. R., Thomas
	P. A., 1998, MNRAS, 301, 478
\bibitem[\protect\citeauthoryear{Thomas \& Couchman}{1992}]{tc92} 
	Thomas P. A., Couchman H. M. P., 1992, MNRAS, 257, 11
\bibitem[\protect\citeauthoryear{Teyssier}{2002}]{tey02} 
	Teyssier R., 2002, A\&A, 385, 337
\bibitem[\protect\citeauthoryear{van den Bosch}{2001}]{vdb01} 
	van den Bosch F. C., 2001, MNRAS, 327, 133
\bibitem[\protect\citeauthoryear{Weil, Eke \& Efstathiou}{1998}]{wee98} 
	Weil M. L, Eke V. R., Efstathiou G., 1998, MNRAS, 300, 773
\bibitem[\protect\citeauthoryear{White \& Rees}{1978}]{wr78} 
	White S. D. M., Rees M. J., 1978, MNRAS, 183, 341
\bibitem[\protect\citeauthoryear{Yepes et al.}{1997}]{yep97} 
	Yepes G., Kates R., Khokhlov A., Klypin A., 1997, MNRAS, 284,
	235
\end{thebibliography}
\end{document}